\documentclass[letterpaper]{JHEP3}
\usepackage[T1]{fontenc}
\usepackage[latin9]{inputenc}
\usepackage{amssymb}
\usepackage{esint}
\usepackage{graphicx}

\makeatletter

\providecommand{\tabularnewline}{\\}

\usepackage{skak}
\title{ Twisted Magnons 
\vspace{-0.6cm}}
\preprint{YITP-SB-10-43}
\author{
Abhijit Gadde$^1$
and Leonardo Rastelli$^2$
\\
\\
\it C.N. Yang Institute for Theoretical Physics,\\
\it Stony Brook University, \\
\it Stony Brook, NY 11794-3840, USA \\
\\
$^1$
\tt{abhijit@insti.physics.sunysb.edu}
\\
$^2$
\tt{leonardo.rastelli@stonybrook.edu}
}
\bigskip

\abstract{
\vspace{0.2cm}

We study spin chains  for superconformal quiver
gauge theories in the moduli space of  ${\cal N}=2$  orbifolds.  
Independent of integrability,
which is generally broken, we use the centrally extended $SU(2|2)$ symmetry 
of the magnons to   fix their dispersion relations and two-body S-matrices,
 as functions of the exactly marginal couplings.   
 }

\renewcommand{\[}{\begin{equation}} 
\renewcommand{\]}{\end{equation}} 

\makeatother

\begin{document}
\global\long\def\gbar{g_{+}}

\global\long\def\gdiff{g_{-}}

\global\long\def\gen#1{\mathcal{#1}}

\global\long\def\QQ{{\cal Q}}

\global\long\def\xp{x^{+}}

\global\long\def\xm{x^{-}}

\global\long\def\xcp{\check{x}^{+}}

\global\long\def\xcm{\check{x}^{-}}

\global\long\def\gp{G}

\global\long\def\sec#1{\tilde{#1}}

\section{Introduction}

The spin chain associated to the planar dilation operator of ${\cal N}=4$ super-Yang Mills \cite{Minahan:2002ve, Beisert:2003yb, Beisert:2003tq} is strongly constrained by symmetry. While the structure of the Hamiltonian becomes unwieldy beyond
one loop, and no closed form  is yet in sight, the S-matrix of  magnon excitations of the infinite chain is
a relatively simple object \cite{Staudacher:2004tk, Beisert:2005tm, Beisert:2005fw}. 
Assuming integrability (for which there is by now strong evidence),
the $n$-body S-matrix factorizes in terms of two-body S-matrices.
In turn, the full matrix structure of the two-body S-matrix is fixed by  Beisert's centrally extended $SU(2|2) \times SU(2|2)$ symmetry \cite{Beisert:2005tm}. 
Finally, the overall phase  is determined with the help of  crossing symmetry and plausible physical assumptions \cite{Janik:2006dc, Beisert:2006ez, Arutyunov:2006iu, Beisert:2006ib}. 
 
The centrally extended $SU(2|2)$ symmetry is a general feature of  spin chains for ${\cal N}=2$ $4d$
superconformal theories\footnote{See also \cite{Agarwal:2010tx} for applications of $SU(2|2)$ to a class theories
with 16 supercharges.}, indeed $SU(2|2)$ is a subgroup of the ${\cal N}=2$ superconformal group $SU(2,2|2)$
preserved by the choice of the spin chain vacuum. In this paper we explore the consequences 
of this symmetry in a class of ${\cal N}=2$ SCFTs, the 
  quiver theories related by exactly marginal deformations to  
${\cal N}=2$ orbifolds of ${\cal N}=4$ super-Yang Mills. 

Unlike the case of
   ${\cal N}=4$ SYM,  only one copy of the $SU(2|2)$ supergroup is preserved,
 while the other is broken to its bosonic subgroup. 
 We show how to fix the dispersion relations and
 two-body  S-matrices of the magnons transforming under the surviving $SU(2|2)$
 by a generalization of Beisert's approach.  Since the $SU(2|2)$ representations
 are  now ``twisted'', the generalization is not entirely trivial and leads to
 interesting functions of the exactly marginal couplings. 
 At the orbifold point  the magnons are gapless and the spin chain is integrable \cite{Beisert:2005he, Solovyov:2007pw} but as we perturb away from it, the magnons
 acquire a gap, and their two-body S-matrices do not satisfy the Yang-Baxter equation.
 So for general values of the couplings
 the theories are not integrable, and  the
  complete magnon S-matrix cannot be deduced from the two-body S-matrix. 
  Nevertheless the  dispersion
  relations and two-body S-matrices are 
  interesting pieces of information in their own right, 
 and it is remarkable that one can obtain for them all-order expressions. 
 At one-loop, we find agreement with the explicit perturbative calculations of
 \cite{Gadde:2010zi, fullH}.
 At strong 't Hooft coupling, one should be able to compare our field-theoretic
 results   with a giant-magnon  \cite{Hofman:2006xt} calculation in the dual string theory, which is a deformation
 of the orbifold background $AdS_5 \times S^5/\Gamma$ \cite{Kachru:1998ys, Klebanov:1999rd}.

For ease of notation, in most of the paper we focus on the simplest case, the ${\cal N}=2$ superconformal quiver with
$SU(N_c) \times SU(N_{\check c})$ gauge group,\footnote{The two gauge groups are identical, $N_c \equiv N_{\check c}$, but
we find it useful to always denote with a ``check'' quantities associated to the second gauge group.}
which is in the moduli space of the $\mathbb{Z}_2$ orbifold of ${\cal N}=4$ SYM.  In section
\ref{dispersions} we determine the dispersion relation of the bifundamental magnons and in section \ref{Smatrix}  their
two-body  S-matrix.

Following Berenstein et al$.$ \cite{Berenstein:2005jq}, in section \ref{emergent} we re-derive
the dispersion relations of the twisted magnons 
from a large $N$ analysis of the quiver matrix model, 
obtained by quantizing the gauge theory on $S^3 \times \mathbb{R}$ and
keeping the zero modes on $S^3$. It is not a priori obvious that this approach, which relies
on an uncontrolled approximation, should give the same answer as the exact algebraic analysis, but it does.
This viewpoint gives  a simple geometric interpretation of dispersion relations, 
very suggestive of an emergent dual geometry.

The generalization
to ${\cal N}=2$ $\mathbb{Z}_k$ orbifolds
is straightforward, and we indicate it in section~\ref{Z_k}.

\medskip

In the rest of this introduction we describe the symmetry structure
of the $\mathbb{Z}_2$-quiver spin chain, contrasting it with the ${\cal N}=4$ chain.
This will serve as an overview of our logic and to orient the reader through our  notations.

The  superconformal symmetry of ${\cal N}=4$ SYM
is $PSU(2,2|4)$. It is broken to $PSU(2|2)\times PSU(2|2)\times\mathbb{R}$,
where $\mathbb{R}$ is a central generator corresponding to the spin chain Hamiltonian,
by the choice of the BMN \cite{Berenstein:2002jq} vacuum ${\rm Tr} \, \Phi^J$. The magnon excitations on
this vacuum are in the fundamental
representation of the unbroken symmetry, and they are gapless because
they are the Goldstone modes associated to the broken generators. 
The $PSU(2,2|4)$ symmetry generators are shown in
table \ref{tab:gen}. The boxed generators, in the diagonal blocks, are preserved
by the choice of the vacuum while the off-diagonal ones are broken and
 correspond to the magnons. The broken generators are labelled in terms of the corresponding magnons: the upper-right  block contains the magnon creation operators  and the lower-left block the magnon annihilation operators.

\begin{table}[h]
\begin{centering}
\begin{tabular}{ccc|cc}
 & $SU(2_{\dot{\alpha}})$ & \multicolumn{1}{c}{$SU(2_{I})$} & $SU(2_{\alpha})$ & $SU(2_{\hat{I}})$\tabularnewline
\cline{2-3} 
\multicolumn{1}{c|}{$SU(2_{\dot{\alpha}})$} & \multicolumn{1}{c|}{$\gen L_{\:\dot{\beta}}^{\dot{\alpha}}$} & $\gen Q_{\: J}^{\dot{\alpha}}$ & $D_{\:\beta}^{\dagger\dot{\alpha}}$ & $\lambda_{\:\hat{J}}^{\dagger\dot{\alpha}}$\tabularnewline
\cline{2-3} 
\multicolumn{1}{c|}{$SU(2_{I})$} & \multicolumn{1}{c|}{$\gen S_{\:\dot{\beta}}^{I}$} & $\gen R_{\: J}^{I}$ & $\lambda_{\:\beta}^{\dagger I}$ & $\mathcal{X}_{\:\hat{J}}^{\dagger I}$\tabularnewline
\cline{2-5} 
$SU(2_{\alpha})$ & $D_{\:\dot{\beta}}^{\alpha}$ & $\lambda_{\: J}^{\alpha}$ & \multicolumn{1}{c|}{$\gen L_{\:\beta}^{\alpha}$} & \multicolumn{1}{c|}{$\gen Q_{\:\hat{J}}^{\alpha}$}\tabularnewline
\cline{4-5} 
$SU(2_{\hat{I}})$ & $\lambda_{\:\dot{\beta}}^{\hat{I}}$ & $\mathcal{X}_{\: J}^{\hat{I}}$ & \multicolumn{1}{c|}{$\gen S_{\:\beta}^{\hat{I}}$} & \multicolumn{1}{c|}{$\gen R_{\:\hat{J}}^{\hat{I}}$}\tabularnewline
\cline{4-5} 
\end{tabular}
\par\end{centering}

\caption{\label{tab:gen}The $PSU(2,2|4)$ symmetry generators. The R-symmetry subgroup 
$SU(4)$ is represented as branched into $SU(2_{I})\times SU(2_{\hat{I}})$.
We have introduced the notation $SU(2_{\alpha})$ for $SU(2)_{\alpha}$ etc.}
\end{table}

A priori, the two-body magnon S-matrix, decomposed according to the $SU(2_{\alpha}|2_{\hat{I}})\times SU(2_{\dot{\alpha}}|2_{I})$ quantum numbers,
can take the schematic form
\[
S_{SU(2_{\alpha}|2_{\hat{I}})\times SU(2_{\dot{\alpha}}|2_{I})}=S_{SU(2_{\alpha}|2_{\hat{I}})}\otimes S_{SU(2_{\dot{\alpha}}|2_{I})}+S_{SU(2_{\alpha}|2_{\hat{I}})}^{\prime}\otimes S_{SU(2_{\dot{\alpha}}|2_{I})}^{\prime}+\ldots\]
As it turns out, the $SU(2|2)$ S-matrix is unique up to an overall phase \cite{Beisert:2005tm}, so
one has the useful factorization  \[
S_{SU(2_{\alpha}|2_{\hat{I}})\times SU(2_{\dot{\alpha}}|2_{I})}=S_{SU(2_{\alpha}|2_{\hat{I}})}\otimes S_{SU(2_{\dot{\alpha}}|2_{I})}\,.
\]
The $SU(2_{\alpha}|2_{\hat{I}})$ S-matrix describes the scattering
of magnons in the highest weight state of $SU(2_{\dot{\alpha}}|2_{I})$,
and viceversa. 

The $\mathbb{Z}_{2}$ 
projection of ${\cal N}=4$ SYM breaks $PSU(2_{\alpha},2_{\dot{\alpha}}|4_{I\hat{I}})$
to $SU(2_{\alpha},2_{\dot{\alpha}}|2_{I})\times SU(2_{\hat{I}})$.  At the orbifold point $g_{YM} = \check g_{YM}$
the breaking  is only  global (by boundary conditions on the periodic chain), but
for general  couplings  the $PSU(2_{\alpha},2_{\dot{\alpha}}|4_{I\hat{I}})$ is truly lost. 
The symmetry preserved by the spin chain vacuum is
 $SU(2_{\dot{\alpha}}|2_{I})\times SU(2_{\alpha})\times SU(2_{\hat{I}})$.
 Table \ref{tab:genN2} lists the symmetry generators of the theory,
 with the broken generators 
 identified as Goldstone modes.
 The Goldstone excitations (gapless magnons) are in the fundamental
representation of $SU(2_\alpha) \times SU(2_{\dot{\alpha}}|2_{I})$. The $\{ \mathcal{X}_{\:\hat{J}}^{I}\,, \lambda_{\:\hat{J}}^{\dot{\alpha}}\}$
magnons, in the fundamental of  $SU(2_{\hat I}) \times SU(2_{\dot{\alpha}}|2_{I})$,  are omitted in table  \ref{tab:genN2} 
because they do not correspond to broken generators -- indeed they  have  a gap for $g_{YM} \neq \check g_{YM}$.
Their dynamics is the main focus of this paper.

Here we are using the ``orbifold'' notation, where the fields are labeled as in ${\cal N}=4$ SYM, and
are  $2N_c \times 2N_c$ matrices in color space (see equ.(\ref{survive})). The state space of the spin chain
consists of an twisted and and untwisted sector, distinguished by whether or not the twist operator $\tau$ (equ.(\ref{tau})) 
is inserted on the chain. The two sectors mix for $g_{YM} \neq \check g_{YM}$. In particular
 the symmetry generators and the central charges acquire twisted components, see~(\ref{twistedgen}, \ref{twistedcentral}).

\begin{table}[h]
\begin{centering}
\begin{tabular}{ccc|cc}
 & $SU(2_{\dot{\alpha}})$ & \multicolumn{1}{c}{$SU(2_{I})$} & $SU(2_{\alpha})$ & $SU(2_{\hat{I}})$\tabularnewline
\cline{2-3} 
\multicolumn{1}{c|}{$SU(2_{\dot{\alpha}})$} & \multicolumn{1}{c|}{$\gen L_{\:\dot{\beta}}^{\dot{\alpha}}$} & $\gen Q_{\: J}^{\dot{\alpha}}$ & $D_{\:\beta}^{\dagger\dot{\alpha}}$ & \tabularnewline
\cline{2-3} 
\multicolumn{1}{c|}{$SU(2_{I})$} & \multicolumn{1}{c|}{$\gen S_{\:\dot{\beta}}^{I}$} & $\gen R_{\: J}^{I}$ & $\lambda_{\:\beta}^{\dagger I}$ & \tabularnewline
\cline{2-4} 
$SU(2_{\alpha})$ & $D_{\:\dot{\beta}}^{\alpha}$ & $\lambda_{\: J}^{\alpha}$ & \multicolumn{1}{c|}{$\gen L_{\:\beta}^{\alpha}$} & \tabularnewline
\cline{4-5} 
$SU(2_{\hat{I}})$ &  & \multicolumn{1}{c}{} & \multicolumn{1}{c|}{} & \multicolumn{1}{c|}{$\gen R_{\:\hat{J}}^{\hat{I}}$}\tabularnewline
\cline{5-5} 
\end{tabular}
\par\end{centering}

\caption{\label{tab:genN2}The  generators of $SU(2,2|2) \times SU(2_{\hat I})$,
the symmetry of the $\mathbb{Z}_2$ quiver.
 As before, the boxed generator are preserved by
the choice of the spin-chain vacuum while the other correspond to Goldstone excitations.}

\end{table}

The scattering of any two magnons (gapless or gapped) is given
by a factorized two-body S-matrix, 
\[
S_{SU(2_{\alpha})\times SU(2_{\hat{I}})\times SU(2_{\dot{\alpha}}|2_{I})}= {S}_{SU(2_{\alpha})\times SU(2_{\hat{I}})}\otimes S_{SU(2_{\dot{\alpha}}|2_{I})}\, .\]
The  $S_{SU(2_{\dot{\alpha}}|2_{I})}$ S-matrix
 describes the scattering of  magnons in the highest weight of $SU(2_{\alpha})\times SU(2_{\hat{I}})$.
It has both an untwisted and a twisted component, schematically
\begin{equation}
S_{SU(2_{\dot{\alpha}}|2_{I})}\, | {\cal X}_1  {\cal X}_2 \rangle  ={\cal S}^{\mathbb{I}}\,  | {\cal X}_1  {\cal X}_2 \rangle +  
{\cal S}^{\tau} \, | {\cal X}_1  {\cal X}_2 \tau \rangle\,.
\end{equation}
The centrally extended $SU(2|2)$ symmetry will fix both components uniquely, up to
the usual phase ambiguity.

\section{Magnon Dispersion Relations}
\label{dispersions}
\subsection{Review:   ${\cal N}=4$  magnons}

The field content of ${\cal N}=4$ super
Yang-Mills consists of the gauge field $A_{\mu}$, four Weyl
spinors $\lambda_{\alpha}^{A}$ and six real scalars $X^{i}$, where
$A=1,\ldots4$ and $i=1,\ldots6$ are indices labelling fundamental
and antisymmetric self-dual representation of the $SU(4_A)$ R-symmetry
group respectively. 
Under $U(1)_{r}\times SU(2_{I})_{R}\times SU(2_{\hat{I}})_{L} \subset SU(4_{A})$, the scalars branch
into one complex scalar $\Phi$, charged under $U(1)_r$,
 and $SU(2_{I})_{R}\times SU(2_{\hat{I}})_{L}$ bifundamental scalars ${\cal X}^{I\hat{I}}$, with zero $U(1)_r$ charge,
satisfying the reality condition ${\cal X}^{I\hat{I}\dagger}=-\epsilon^{IJ}\epsilon^{\hat{I}\hat{J}}{\cal X}^{J\hat{J}}$.
The fermions decompose as $\lambda_{\alpha}^{I}$ and $\lambda_{\alpha}^{\hat{I}}$. The
${\cal N}=2$ supersymmetry organizes $A_{\mu},\lambda_{\alpha}^{I},\Phi$
into a vector multiplet and ${\cal X}^{I\hat{I}},\lambda_{\alpha}^{\hat{I}}$
into a hypermultiplet.

\label{sub:N=00003D4disp}

For definiteness we focus on the ``right-handed'' magnons,
in the fundamental of
$SU(2_{\dot{\alpha}}|2_{I})$  and in the
 highest-weight state of
of $SU(2_{\alpha}|2_{\hat{I}})$,
\begin{equation}
\mathcal{X}_{\:\hat{+}}^{I}\equiv\mathcal{X}^{I} \, ,\quad
\lambda_{\:\hat{+}}^{\dot{\alpha}}\equiv\lambda^{\dot{\alpha}}\,.
\end{equation}
Beisert determined the magnon dispersion relation from symmetry arguments, as we now review.
The non-zero commutation relations of the $SU(2|2)$ generators are:\begin{eqnarray*}
[\gen R_{\: J}^{I},\gen J^{K}] & = & \delta_{J}^{K}\gen J^{I}-\frac{1}{2}\delta_{J}^{I}\gen J^{K}\\
{}[\gen L_{\:\dot{\beta}}^{\dot{\alpha}},\gen J^{\dot{\gamma}}] & = & \delta_{\:\dot{\beta}}^{\dot{\gamma}}\gen J^{\dot{\alpha}}-\frac{1}{2}\delta_{\dot{\beta}}^{\dot{\alpha}}\gen J^{\dot{\gamma}} \\
\{\gen Q_{\: I}^{\dot{\alpha}},\gen S_{\:\dot{\beta}}^{J}\} & = & \delta_{I}^{J}\gen L_{\:\dot{\beta}}^{\dot{\alpha}}+\delta_{\dot{\beta}}^{\dot{\alpha}}\gen R_{\: I}^{J}+\delta_{I}^{J}\delta_{\dot{\beta}}^{\dot{\alpha}}\gen C
\end{eqnarray*}
where $\gen J$ represents any generator with the appropriate index.
The central
charge $\gen C$ is related to the scaling dimension as $\gen C=\frac{1}{2}(\Delta-|r|)$.
The impurities $({\cal X}^{\cal I}\,,\lambda^{\dot \alpha})$ transform in the  fundamental representation 
of $SU(2|2)$, and closure of the algebra fixes $\gen C =\frac{1}{2}$, corresponding to the canonical
dimensions $\Delta=1$ and $\Delta=\frac{3}{2}$ for ${\cal X}$ and $\lambda$.
Consider now a magnon of momentum $p$,
\[
\Psi(p)=\sum_{l=-\infty}^{\infty}e^{ipl}|\mathcal{X}(l) \, \rangle .\]
For $p \neq 0$, the state acquires a non-vanishing anomalous dimension, so $\gen C\neq \frac{1}{2}$,
but the representation remains short,
as there are no other degrees of freedom with which it could combine to become long.
This is in conflict with the $SU(2|2)$ algebra. The resolution is to allow for a further
central extension by momentum-dependent central charges 
$\gen P$ and $\gen K$,
\[
\{\gen Q_{\: I}^{\dot{\alpha}},\gen Q_{\: J}^{\dot{\beta}}\}=\epsilon^{\dot{\alpha}\dot{\beta}}\epsilon_{IJ}\gen P,\quad\{\gen S_{\:\dot{\alpha}}^{I},\gen S_{\:\dot{\beta}}^{J}\}=\epsilon^{IJ}\epsilon_{\dot{\alpha}\dot{\beta}}\gen K \,  . \]
The most general action of the generators in the excitation picture is
:\begin{eqnarray}
\gen Q_{\: I}^{\dot{\alpha}}|\mathcal{X}^{J}\rangle & = & a\delta_{I}^{J}|\lambda^{\dot{\alpha}}\rangle \\
\gen Q_{\: I}^{\dot{\alpha}}|\lambda^{\dot{\beta}}\rangle & = & b\epsilon^{\dot{\alpha}\dot{\beta}}\epsilon_{IJ}|\mathcal{X}^{J}\Phi^{+}\rangle \nonumber\\
\gen S_{\:\dot{\alpha}}^{I}|\mathcal{X}^{J}\rangle & = & c\epsilon^{IJ}\epsilon_{\dot{\alpha}\dot{\beta}}|\lambda^{\dot{\beta}}\Phi^{-}\rangle \nonumber\\
\gen S_{\:\dot{\alpha}}^{I}|\lambda^{\dot{\beta}}\rangle & = & d\delta_{\dot{\alpha}}^{\dot{\beta}}|\mathcal{X}^{I}\rangle\,\nonumber ,
\end{eqnarray}
which implies 
\begin{eqnarray}
\gen{P|}\mathcal{X}\rangle & = & ab|\mathcal{X}\Phi^{+}\rangle\label{eq:genP}\\
\gen{K|}\mathcal{X}\rangle & = & cd|\mathcal{X}\Phi^{-}\rangle\,.\label{eq:genK}\\
\gen{C }|\mathcal{X}\rangle & = &  \frac{1}{2}(ad+bc) |\mathcal{X}\rangle\,. \label{eq:genC}
\end{eqnarray}
Closure of the algebra requires $ad-bc=1$. We can then formally solve
\[ \label{formally}
\gen C  =  \frac{1}{2}\sqrt{1+4\gen P \gen K} \,.
\]

For a quick heuristic derivation of the central charges, we can proceed
as follows. The supersymmetry transformations of
the fields appearing in the Lagrangian,\begin{eqnarray*}
\gen Q_{\: I}^{\dot{\alpha}}\mathcal{X}^{K} & = & \delta_{I}^{K}\lambda^{\dot{\alpha}}\\
\gen Q_{\: J}^{\dot{\beta}}\lambda^{\dot{\alpha}} & = & \epsilon^{\dot{\beta}\dot{\alpha}}\frac{\partial W}{\partial\mathcal{X}^{J}}=\frac{g}{\sqrt{2}}\epsilon^{\dot{\beta}\dot{\alpha}}\epsilon_{JL}[\mathcal{X}^{L},\Phi]\end{eqnarray*}
where $W=\frac{g}{\sqrt{2}}\mbox{Tr }\mathcal{X}^{I\hat{I}}\Phi\mathcal{X}_{I\hat{I}}$
is the superpotential of ${\cal N}=4$ super Yang-Mills. 
The coupling $g$ is the square root of the 't Hooft coupling,
normalized as
\begin{equation}
g^2 = \frac{g_{YM}^2 N_c}{8 \pi^2}\,.
\end{equation}
These susy transformations
lead to the anticommutators
\begin{eqnarray*}
\{\gen Q_{\: I}^{\dot{\alpha}},\gen Q_{\: J}^{\dot{\beta}}\}\mathcal{X}^{K} & = & \frac{g}{\sqrt{2}}\epsilon^{\dot{\alpha}\dot{\beta}}\epsilon_{IJ}[\Phi,\mathcal{X}^{K}]\\
\{\gen Q_{\: I}^{\dot{\alpha}},\gen Q_{\: J}^{\dot{\beta}}\}\lambda^{\dot{\gamma}} & = & \frac{g}{\sqrt{2}}\epsilon^{\dot{\alpha}\dot{\beta}}\epsilon_{IJ}[\Phi,\lambda^{\dot{\gamma}}]\end{eqnarray*}
Using the fact that momentum eigenstates
 satisfy\begin{equation}
|\Phi^{\pm}\mathcal{X}\rangle=e^{\mp ip}|\mathcal{X}\Phi^{\pm}\rangle\, ,\label{eq:through}\end{equation}
we can realize the susy transformation laws
on the spin chain as
\[
\{\gen Q_{\: I}^{\dot{\alpha}},\gen Q_{\: J}^{\dot{\beta}}\}|\mathcal{X}\rangle=\epsilon^{\dot{\alpha}\dot{\beta}}\epsilon_{IJ}\gen P|\mathcal{X}\rangle=\epsilon^{\dot{\alpha}\dot{\beta}}\epsilon_{IJ}\frac{g}{\sqrt{2}}(e^{-ip}-1)|\mathcal{X}\Phi^{+}\rangle\, ,\]
implying $ab=\frac{g}{\sqrt{2}}(e^{-ip}-1)$. Similarly using
$\{\gen S,\gen S\}$, we can obtain $cd=\frac{g}{\sqrt{2}}(e^{ip}-1)$.
 Finally, from (\ref{formally}),
\[
\label{dispersionN4}
\Delta-|r|  = 2 \gen C =   \sqrt{1+8g^{2}\sin^{2}\frac{p}{2}}\, .
\]
This derivation\footnote{The first field-theoretic
argument for  the square-root form (\ref{dispersionN4})  was  given in \cite{Santambrogio:2002sb}.}  
is only heuristic because of the assumption that the susy transformations
in the excitation picture can be simply read off from the classical Lagrangian.
In \cite{Beisert:2005tm}, Beisert used a purely algebraic method to determine
the central charges, as we review in appendix
\ref{sec:Beisertmethod}. The algebraic method confirms
the form  (\ref{dispersionN4}), but with $g^2$  a priori replaced by a renormalized coupling ${\bf g}^2 = g^2 + O(g^4)$.
There is strong evidence that in ${\cal N}=4$ SYM ${\bf g}^2 = g^2$. In the ABJM theory \cite{Aharony:2008ug}
one can run an identical  argument, but the coupling {\it is} renormalized \cite{Nishioka:2008gz, Gaiotto:2008cg, Grignani:2008is}.
See \cite{Bak:2009mq, Berenstein:2009qd}
for  discussions of this issue.

\subsection{The  $\mathbb{Z}_{2}$ orbifold and its deformation}

The $\mathbb{Z}_{2}$ orbifold theory is the well known quiver gauge
theory living on the worldvolume of D3 branes probing $\mathbb{R}^{2}\times\mathbb{R}^{4}/\mathbb{Z}_{2}$
singularity. It is obtained from ${\cal N}=4$ super Yang-Mills by
projecting onto the $\mathbb{Z}_{2}\subset SU(2)_{L}$ invariant states.
The $\mathbb{Z}_{2}$ action identifies $\mathcal{X}^{I\hat{I}}\to-{\cal X}^{I\hat{I}}$
while acting trivially on $\Phi$. The supersymmetry is broken to
${\cal N}=2$ as the supercharges with $SU(2)_{L}$ indices are projected
out. The $SU(4)$ R symmetry group is broken to $SU(2)_{R}\times SO(3)_{L}\times U(1)_{r}$.
$SU(2)_{R}\times U(1)_{r}$ is the R symmetry group of the ${\cal N}=2$
theory while $SO(3)_{L}$ is a global symmetry.
In  color space, we start with  $SU(2N_{c})$ gauge group and
declare the nontrivial element of the orbifold to be \[
\label{tau}
\tau=\left(\begin{array}{cc}
\mathbb{I}_{N_{c}\times N_{c}} & 0\\
0 & -\mathbb{I}_{N_{\check{c}}\times N_{\check{c}}}\end{array}\right).\]
It acts on the fields of ${\cal N}=4$ SYM as\[
A_{\mu}\to\tau A_{\mu}\tau,\qquad\Phi\to\tau\Phi\tau,\qquad\lambda^{I}\to\tau\lambda^{I}\tau,\qquad{\cal X}^{I\hat{I}}\to-\tau{\cal X}^{I\hat{I}}\tau,\qquad\lambda^{\hat{I}}\to-\tau\lambda^{\hat{I}}\tau.\]
The components that survive the projection are\begin{eqnarray}
\label{survive}
A_{\mu} & = & \left(\begin{array}{cc}
A_{\mu} & 0
\\
0 & \check{A}_{\mu}\end{array}\right),\quad\Phi=\left(\begin{array}{cc}
\phi & 0\\
0 & \check{\phi}\end{array}\right),\quad\lambda^{I}=\left(\begin{array}{cc}
\lambda^{I} & 0\\
0 & \check{\lambda}^{I}\end{array}\right),\\
{\cal X}^{I\hat{I}} & = & \left(\begin{array}{cc}
0 & Q^{I\hat{I}}\\
\bar{Q}^{I\hat{I}} & 0\end{array}\right),\quad\lambda^{\hat{I}}=\left(\begin{array}{cc}
0 & \psi^{\hat{I}}\\
\tilde{\psi}^{\hat{I}} & 0\end{array}\right). \nonumber \end{eqnarray}
The orbifold theory has an untwisted sector of states,
which descend by projection from ${\cal N}=4$, and a twisted sector
of states, characterized by the presence of one insertion of the twist
operator $\tau$ in the color trace. We refer to this presentation of the theory (in terms of $2N_c \times 2 N_c$ matrices)
as the ``orbifold basis''.

Equivalently, we can present the theory as an ${\cal N}=2$ quiver
gauge theory with  product gauge group
$SU(N_{c})\times SU(N_{\check{c}})$ and two bifundamental hypermultiplets:
$(A_{\mu},\lambda^{I},\phi)$
and $(\check{A}_{\mu},\check{\lambda},\check{\phi})$ are the two
vector multiplets while $(Q^{I\hat{I}},\psi^{\hat{I}})$ and $(\bar{Q}^{I\hat{I}},\tilde{\psi}^{\hat{I}})$
are the two hypermultiplets transforming respectively  in the ${\bf N}_{c}\times\overline{{\bf N}}_{\check{c}}$
and $\overline{{\bf N}}_{c}\times{\bf N}_{\check{c}}$ representations.

The two gauge couplings  $g$ and $\check{g}$ are exactly
marginal. For $g \neq \check g$ the  superpotential acquires a twisted term,
\[ \label{deformedW}
W=\frac{\gp}{\sqrt{2}}\mbox{Tr }[\frac{1}{2}(\sqrt{\kappa}+\frac{1}{\sqrt{\kappa}})+\tau\frac{1}{2}(\sqrt{\kappa}-\frac{1}{\sqrt{\kappa}})]\mathcal{X}^{I\hat{I}}\Phi\mathcal{X}_{I\hat{I}}\]
where
\[
\gp \equiv \sqrt{g\check{g}}\,, \qquad
\kappa \equiv \frac{\check{g} }{g} \,.
\]
 In the quiver language,
\begin{eqnarray*}
W & = & \frac{g}{\sqrt{2}}\mbox{Tr }\bar{Q}^{I\hat{I}}\phi Q_{I\hat{I}}+\frac{\check{g}}{\sqrt{2}}Q^{I\hat{I}}\check{\phi}\bar{Q}_{I\hat{I}}\\
 & = & \frac{\gp}{\sqrt{2}}(\mbox{Tr }\frac{1}{\sqrt{\kappa}}\bar{Q}^{I\hat{I}}\phi Q_{I\hat{I}}+\sqrt{\kappa}Q^{I\hat{I}}\check{\phi}\bar{Q}_{I\hat{I}}) \,.\end{eqnarray*}

\subsection{Twisted magnons}

As we have explained in the introduction, the magnons of the $\mathbb{Z}_2$ theory
fall into two classes: Goldstone magnons associated with the broken generators, carrying
an $\alpha$ index, and magnons not associated with symmetries, carrying a $\hat I$ index.
 Both types are in the fundamental representation of $SU(2_{\dot \alpha}|2_I)$.
 The algebraic analysis for the Goldstone magnons is
 exactly as in ${\cal N}=4$ SYM, so they obey the same dispersion relation.
On the other hand, the non-Goldstone  magnons  transform in a {}``twisted'' representation of the $SU(2|2)$
superalgebra, \begin{eqnarray}
\label{twistedgen}
\gen Q_{\: I}^{\dot{\alpha}}|\mathcal{X}^{J}\rangle & = & a_{0}\delta_{I}^{J}|\lambda^{\dot{\alpha}}\rangle+a_{1}\delta_{I}^{J}|\tau\lambda^{\dot{\alpha}}\rangle\\
\gen Q_{\: I}^{\dot{\alpha}}|\lambda^{\dot{\beta}}\rangle & = & b_{0}\epsilon^{\dot{\alpha}\dot{\beta}}\epsilon_{IJ}|\mathcal{X}^{J}\Phi^{+}\rangle+b_{1}\epsilon^{\dot{\alpha}\dot{\beta}}\epsilon_{IJ}|\tau\mathcal{X}^{J}\Phi^{+}\rangle \nonumber \\
\gen S_{\:\dot{\alpha}}^{I}|\mathcal{X}^{J}\rangle & = & c_{0}\epsilon^{IJ}\epsilon_{\dot{\alpha}\dot{\beta}}|\lambda^{\dot{\beta}}\Phi^{-}\rangle+c_{1}\epsilon^{IJ}\epsilon_{\dot{\alpha}\dot{\beta}}|\tau\lambda^{\dot{\beta}}\Phi^{-}\rangle \nonumber \\
\gen S_{\:\dot{\alpha}}^{I}|\lambda^{\dot{\beta}}\rangle & = & d_{0}\delta_{\dot{\alpha}}^{\dot{\beta}}|\mathcal{X}^{I}\rangle+d_{1}\delta_{\dot{\alpha}}^{\dot{\beta}}|\tau\mathcal{X}^{I}\rangle\nonumber
\end{eqnarray}
One then finds for the central charges:
\begin{eqnarray}
\label{twistedcentral}
\gen P|\mathcal{X}\rangle & = & (a_{0}b_{0}+a_{1}b_{1})|\mathcal{X}\Phi^{+}\rangle+(a_{0}b_{1}+a_{1}b_{0})|\tau\mathcal{X}\Phi^{+}\rangle
 \\
\gen K|\mathcal{X}\rangle & = & (c_{0}d_{0}+c_{1}d_{1})|\mathcal{X}\Phi^{-}\rangle+(c_{0}d_{1}+c_{1}d_{0})|\tau\mathcal{X}\Phi^{-}\rangle
\nonumber \\
\gen C|{\cal X}\rangle & = & [\frac{1}{2}(a_{0}d_{0}+b_{0}c_{0})+\frac{1}{2}(a_{1}d_{1}+b_{1}c_{1})]|{\cal X}\rangle\nonumber
\\
 & + & [\frac{1}{2}(a_{0}d_{1}+b_{0}c_{1})+\frac{1}{2}(a_{1}d_{0}+b_{1}c_{0})]|\tau{\cal X}\rangle\,.\nonumber
 \end{eqnarray}
Using the supersymmetry transformations following from the deformed superpotential (\ref{deformedW}),
a little calculation gives
\begin{eqnarray}
a_{0}b_{0}+a_{1}b_{1} & = & \frac{G}{\sqrt{2}}\frac{1}{2}(\frac{1}{\sqrt{\kappa}}+\sqrt{\kappa})(e^{-ip}-1)\\
a_{0}b_{1}+a_{1}b_{0} & = & \frac{G}{\sqrt{2}}\frac{1}{2}(\frac{1}{\sqrt{\kappa}}-\sqrt{\kappa})(e^{-ip}+1) \nonumber
\\
c_{0}d_{0}+c_{1}d_{1} & = & \frac{G}{\sqrt{2}}\frac{1}{2}(\frac{1}{\sqrt{\kappa}}+\sqrt{\kappa})(e^{ip}-1) \nonumber \\
c_{0}d_{1}+c_{1}d_{0} & = & \frac{G}{\sqrt{2}}\frac{1}{2}(\frac{1}{\sqrt{\kappa}}-\sqrt{\kappa})(e^{ip}+1) \,. \nonumber
\end{eqnarray}
We can then read off the central charges
\begin{eqnarray*}
C_{0} & \equiv & \frac{1}{2}(a_{0}d_{0}+b_{0}c_{0})+\frac{1}{2}(a_{1}d_{1}+b_{1}c_{1})=\frac{1}{2}\sqrt{1+8G^{2}\left(\sin^{2}\frac{p}{2}+\frac{1}{4}(\sqrt{\kappa}-\frac{1}{\sqrt{\kappa}})^{2}\right)}\\
C_{1} & \equiv & \frac{1}{2}(a_{0}d_{1}+b_{0}c_{1})+\frac{1}{2}(a_{1}d_{0}+b_{1}c_{0})=0\, .\end{eqnarray*}

It is illuminating to repeat the exercise in the quiver basis, as it
will give us the dispersion relation of the perhaps more ``physical''  bifundamental
excitations that interpolate between the ${\rm Tr}\,  \phi^J$ and ${\rm Tr} \check \phi^J$ vacua.
In the quiver basis, the $({\cal X}, \, \lambda)$ doublet splits
into two doublets, $(Q,\, \psi)$ and $(\bar Q, \, \tilde \psi)$. Let us call these
two fundamental $SU(2|2)$  representations   $V$ and $\sec V$. The action of the algebra ${\cal A}:V\to V$
and ${\cal A}:\sec V\to\sec V$ is given in table \ref{tab:quiverrep}.
\begin{table}[h!]
\begin{centering}
$\begin{array}{ccccccc}
\gen Q_{\: I}^{\dot{\alpha}}|Q^{J}\rangle & = & a\delta_{I}^{J}|\psi^{\dot{\alpha}}\rangle\qquad\qquad & \qquad & \gen Q_{\: I}^{\dot{\alpha}}|\bar{Q}^{J}\rangle & = & \sec a\delta_{I}^{J}|\tilde{\psi}^{\dot{\alpha}}\rangle\qquad\qquad\\
\gen Q_{\: I}^{\dot{\alpha}}|\psi^{\dot{\beta}}\rangle & = & b\epsilon^{\dot{\alpha}\dot{\beta}}\epsilon_{IJ}|Q^{J}\check{\phi}^{+}\rangle\quad & \qquad & \gen Q_{\: I}^{\dot{\alpha}}|\tilde{\psi}^{\dot{\beta}}\rangle & = & \sec b\epsilon^{\dot{\alpha}\dot{\beta}}\epsilon_{IJ}|\bar{Q}^{J}\phi^{+}\rangle\quad\\
\gen S_{\:\dot{\alpha}}^{I}|Q^{J}\rangle & = & c\epsilon^{IJ}\epsilon_{\dot{\alpha}\dot{\beta}}|\psi^{\dot{\beta}}\check{\phi}^{-}\rangle\quad & \qquad & \gen S_{\:\dot{\alpha}}^{I}|\bar{Q}^{J}\rangle & = & \sec c\epsilon^{IJ}\epsilon_{\dot{\alpha}\dot{\beta}}|\tilde{\psi}^{\dot{\beta}}\phi^{-}\rangle\quad\\
\gen S_{\:\dot{\alpha}}^{I}|\psi^{\dot{\beta}}\rangle & = & d\delta_{\dot{\alpha}}^{\dot{\beta}}|Q^{I}\rangle\qquad\qquad & \qquad & \gen S_{\:\dot{\alpha}}^{I}|\tilde{\psi}^{\dot{\beta}}\rangle & = & \sec d\delta_{\dot{\alpha}}^{\dot{\beta}}|\bar{Q}^{I}\rangle.\qquad\qquad\end{array}$
\par\end{centering}

\caption{\label{tab:quiverrep}Representation of the magnons in the quiver
basis.}

\end{table}

The $a,b,c,d$ coefficients in this basis are related to the coefficients
in the orbifold basis
as $a=a_{0}+a_{1}$, $\sec a=a_{0}-a_{1}$ and so
on.  One easily finds
\begin{eqnarray} \label{centralcharges}
ab & = & \frac{G}{\sqrt{2}}(\frac{e^{-ip}}{\sqrt{\kappa}}-\sqrt{\kappa})\equiv P\qquad\qquad\sec a\sec b=\frac{G}{\sqrt{2}}(e^{-ip}\sqrt{\kappa}-\frac{1}{\sqrt{\kappa}})\equiv\sec P\\
cd & = & \frac{G}{\sqrt{2}}(\frac{e^{+ip}}{\sqrt{\kappa}}-\sqrt{\kappa})\equiv K\qquad\qquad\sec c\sec d=\frac{G}{\sqrt{2}}(e^{+ip}\sqrt{\kappa}-\frac{1}{\sqrt{\kappa}})\equiv\sec K\,\nonumber.
\end{eqnarray}
Finally the dispersion relations for $(Q, \psi)$ and
 $(\bar{Q}. \tilde \psi)$ are

\begin{eqnarray}
\Delta - |r| = 2 C & = & \sqrt{1+4PK}=\sqrt{1+8G^{2}\left(\sin^{2}\frac{p}{2}+\frac{1}{4}(\sqrt{\kappa}-\frac{1}{\sqrt{\kappa}})^{2}\right)}\\
\sec \Delta - |r| = 2\sec C & = & \sqrt{1+4\check{P}\check{K}}=\sqrt{1+8G^{2}\left(\sin^{2}\frac{p}{2}+\frac{1}{4}(\sqrt{\kappa}-\frac{1}{\sqrt{\kappa}})^{2}\right)}\, .
\end{eqnarray}
Recall the definitions $G \equiv \sqrt{g \check g}$, $\kappa \equiv \check g/g$. 
As expected, the non-Goldstone magnons acquire a gap for $g \neq \check g$. The derivation
of the dispersion relation just presented suffers
 from the same criticism as the derivation in the ${\cal N}=4$ case:
a priori we should allow for renormalization of the gauge couplings. 
A purely algebraic method for determining  $\gen P$ and
$\gen K$, along the lines of \cite{Beisert:2005tm}, is described
in the appendix \ref{sec:Beisertmethod}, and confirms this expectation. From symmetry alone, one can only
conclude that both dispersion relations take the form
\begin{equation}\label{renormalized}
2 C = 2 \check C =   \sqrt{1+2 ({\bf g}-{{\bf \check g} })^2+8 {\bf g \check g} \sin^{2}\frac{p}{2}}
\end{equation}
where ${\bf g}(g, \check g) = g +\dots$ and ${\bf \check g}(g, \check g)= \check g + \dots$
are a priori renormalized couplings. (Of course such renormalization is known
to {\it not} occur at the orbifold point $g = \check g$.)
This issue also affects the forthcoming expressions
for the S-matrix: the couplings  $g$ and $\check g$ could in principle be replaced by
$\bf g$ and $\bf \check g$. The expansion of  (\ref{renormalized}) 
agrees at one-loop  with the result of \cite{Gadde:2010zi}. It will be interesting
to test it at higher orders.

\section{Two-body S-matrix}
\label{Smatrix}

The scattering problem is formulated on the infinite spin chain.
The scattering of two Goldstone magnons is uninteresting,
since the matrix structure of their two-body S-matrix is exactly as in ${\cal N}=4$ SYM. 
We will focus on the scattering of two ``non-Goldstone'' magnons,
both in the highest weight of $SU(2_{\hat I})$. 
The scattering of a Goldstone and a non-Goldstone magnon is also non-trivial,
and could be studied by  the same  methods.

 In the quiver basis, because of the index structure of the impurities,
one of the non-Goldstone magnons must be from the $Q$ multiplet and the other from
the $\bar{Q}$ multiplet. Their ordering is fixed, we can have $Q$
type magnons always on left of $\bar{Q}$ type ones, or viceversa.
The scattering is pure reflection. For the case of $Q$ type magnon
on the left of $\bar{Q}$ type magnon, the schematic asymptotic form
of the two body wavefunction is\[
\sum_{x_{1}\ll x_{2}}(e^{ip_{1}x_{1}+ip_{2}x_{2}}+S(p_{2},p_{1})e^{ip_{2}x_{1}+ip_{1}x_{2}})|\ldots\phi Q(x_{1})\check{\phi}\ldots\check{\phi}\bar{Q}(x_{2})\phi\ldots\rangle.\]
This is the definition of the two body S matrix $S(p_{1},p_{2})$.
We dropped the $SU(2|2)$ indices of the excitations for clarity.
Similarly, for the other case where $Q$ is on the right side of $\bar{Q}$,
the aymptotic form of the wavefunction is\[
\sum_{x_{1}\ll x_{2}}(e^{ip_{1}x_{1}+ip_{2}x_{2}}+\check{S}(p_{2},p_{1})e^{ip_{2}x_{1}+ip_{1}x_{2}})|\ldots\check{\phi}\bar{Q}(x_{1})\phi\ldots\phi Q(x_{2})\check{\phi}\ldots\rangle\]
which defines $\check{S}$. The two-body S matrices $S$ and $\check{S}$
are related by exchanging $g\leftrightarrow\check{g}$,\[
S(p_{1},p_{2};g,\check{g})=\check{S}(p_{1},p_{2};\check{g},g).\]
For this reason, without loss of generality, we restrict our analysis
to finding $S(p_{1},p_{2})$.

\subsection{Rapidity variables}

Following Beisert, a preliminary step is to solve for the coefficients
$a,b,c,d$
and $\sec a,\sec b,\sec c,\sec d$ appearing in the magnon representation (table \ref{tab:quiverrep})
in terms of convenient rapidity variables. 

For the representation coefficients of the $Q$ multiplet, we write
\[
a=\gamma,\qquad b=-\frac{G}{\sqrt{2}}\frac{1}{\gamma x^{+}}(x^{+}\sqrt{\kappa}-\frac{x^{-}}{\sqrt{\kappa}}),\qquad c=\frac{G}{\sqrt{2}}\frac{i\gamma'}{x^{-}},\qquad d=-\frac{i}{\gamma'}(\frac{x^{+}}{\sqrt{\kappa}}-x^{-}\sqrt{\kappa})\]
The relative factor between
$\gamma$ and $\gamma^{\prime}$ corresponds to relative rescalings
of the fields $Q$ and $\psi$ and affects the S matrix
as an overall phase.$ $ We choose $\gamma=\gamma^{\prime}$.

For the $\bar{Q}$  coefficients, we write \[
\sec a=\sec{\gamma},\qquad\sec b=-\frac{G}{\sqrt{2}}\frac{1}{\sec{\gamma}\sec x^{+}}(\frac{\sec x^{+}}{\sqrt{\kappa}}-\sec x^{-}\sqrt{\kappa}),\qquad\sec c=\frac{G}{\sqrt{2}}\frac{i\sec{\gamma}}{\sec x^{-}},\qquad\sec d=-\frac{i}{\sec{\gamma}}(\sec x^{+}\sqrt{\kappa}-\frac{\sec x^{-}}{\sqrt{\kappa}}).\]
Both pairs of rapidity variables obey  $\frac{x^{+}}{x^{-}}=\frac{\sec x^{+}}{\sec x^{-}}=e^{ip}$.
For  hermitian representations we have to choose\[
|\gamma|=|i(x^{-}\sqrt{\kappa}-\frac{x^{+}}{\sqrt{\kappa}})|^{1/2},\qquad|\sec{\gamma}|=|i(\frac{\sec x^{-}}{\sqrt{\kappa}}-\sec x^{+}\sqrt{\kappa})|^{1/2}.\]
The closure of the algebra requires $ad-bc=1$ and $\sec a\sec d-\sec b\sec c=1$
{\it i.e.}\begin{eqnarray*}
\frac{x^{+}}{\sqrt{\kappa}}-x^{-}\sqrt{\kappa}+\frac{G^{2}}{2}(\frac{1}{x^{+}\sqrt{\kappa}}-\frac{\sqrt{\kappa}}{x^{-}}) & = & i\\
\sec x^{+}\sqrt{\kappa}-\frac{\sec x^{-}}{\sqrt{\kappa}}+\frac{G^{2}}{2}(\frac{\sqrt{\kappa}}{\sec x^{+}}-\frac{1}{\sec x^{-}\sqrt{\kappa}}) & = & i.\end{eqnarray*}
The central charges are then\begin{eqnarray*}
\gen C & = & \frac{1}{2}+i\frac{G^{2}}{2}(\frac{1}{x^{+}\sqrt{\kappa}}-\frac{\sqrt{\kappa}}{x^{-}})=-i\frac{x^{+}}{\sqrt{\kappa}}+ix^{-}\sqrt{\kappa}-\frac{1}{2}\\
\sec {\gen C} & = & \frac{1}{2}+i\frac{G^{2}}{2}(\frac{\sqrt{\kappa}}{\sec x^{+}}-\frac{1}{\sec x^{-}\sqrt{\kappa}})=-i\sec x^{+}\sqrt{\kappa}+i\frac{\sec x^{-}}{\sqrt{\kappa}}-\frac{1}{2}.\end{eqnarray*}
Although  the expressions for the central charges (=anomalous dimensions) of $Q$ and $\bar{Q}$ 
look different in terms of rapidity variables $x$ and $\sec x$, they are in fact
 equal (by construction) as functions of the  momenta. 

\subsection{The S-matrix}

The S-matrix $S$ is an operator 
\begin{equation}
 S\, : \quad V\otimes\sec V\to V\otimes\sec V
 \end{equation}
  and similarly
  \begin{equation}
 \check S\, : \quad \sec V\otimes  V\to \sec V\otimes V \,.
 \end{equation}
The $SU(2|2)$ algebra  acts on
$V\otimes\tilde{V}$ as follows,
\[
{\cal {\cal A}}(v\times\tilde{v})=({\cal A}v)\times\tilde{v}+(-1)^{F_{{\cal A}}F_{v}}v\times({\cal A}\tilde{v}) \, ,\]
where ${\cal A}$ is an element of the algebra, $v,\tilde{v}$
 vectors in $V$ and $\tilde{V}$, and $F$ the fermion
number. To guarantee the $SU(2|2)$ symmetry of the S-matrix
we simply need to impose the matrix equation $[{\cal A},S]=0$. This is sufficient
to determine $S$ up to an overall phase. 

Following \cite{Beisert:2005tm},
we parametrize the S-matrix as\begin{eqnarray}
S|Q_{1}^{I}\bar{Q}_{2}^{J}\rangle & = & A|Q_{2}^{\{I}\bar{Q}_{1}^{J\}}\rangle+B|Q_{2}^{[I}\bar{Q}_{1}^{J]}\rangle+\frac{1}{2}C\epsilon^{IJ}\epsilon_{\dot{\alpha}\dot{\beta}}|\psi_{2}^{\dot{\alpha}}\tilde{\psi}_{1}^{\dot{\beta}}\phi^{-}\rangle\nonumber \\
S|\psi_{1}^{\dot{\alpha}}\tilde{\psi}_{2}^{\dot{\beta}}\rangle & = & D|\psi_{2}^{\{\dot{\alpha}}\tilde{\psi}_{1}^{\dot{\beta}\}}\rangle+E|\psi_{2}^{[\dot{\alpha}}\tilde{\psi}_{1}^{\dot{\beta}]}\rangle+\frac{1}{2}F\epsilon^{\dot{\alpha}\dot{\beta}}\epsilon_{IJ}|Q_{2}^{I}\bar{Q}_{1}^{J}\phi^{+}\rangle\nonumber \\
S|Q_{1}^{I}\tilde{\psi}_{2}^{\dot{\beta}}\rangle & = & G|\psi_{2}^{\dot{\beta}}\bar{Q}_{1}^{I}\rangle+H|Q_{2}^{I}\tilde{\psi}_{1}^{\dot{\beta}}\rangle\nonumber \\
S|\psi_{1}^{\dot{\alpha}}\bar{Q}_{2}^{J}\rangle & = & K|\psi_{2}^{\dot{\alpha}}\bar{Q}_{1}^{J}\rangle+L|Q_{2}^{J}\tilde{\psi}_{1}^{\dot{\alpha}}\rangle.\end{eqnarray}
The linear constraints obeyed by the S-matrix are listed in equ.(\ref{eqset}). Below we give the solution for the components $A,\, B,\, C,\, G,\,H,\,K,\,L$.  The solution for
$B,\, D$ and $E$ involve lengthier expressions -- they can be readily obtained from equ.(\ref{eqset}) with {\tt Mathematica}'s help.
\begin{eqnarray}
A & = & \frac{\sec x_{1}^{-}x_{2}^{-}}{x_{1}^{-}\sec x_{2}^{-}}(\frac{\sec x_{2}^{+}-x_{1}^{-}}{x_{2}^{-}-\sec x_{1}^{+}})\label{eq:solution}\\
B & = & \sec x_{1}^{-}x_{2}^{-}[\sec x_{1}^{+}x_{2}^{+}\kappa(2x_{2}^{-}x_{1}^{+}\sec x_{2}^{+}-\sec x_{1}^{+}x_{2}^{+}(x_{1}^{-}+\sec x_{2}^{+}))\nonumber \\
 &  & +\sec x_{1}^{-}(2\sec x_{1}^{+}x_{2}^{+}(x_{1}^{-}x_{2}^{+}+\sec x_{2}^{+}(x_{2}^{+}-x_{1}^{-}))\nonumber \\
 &  & +x_{2}^{-}(-2x_{1}^{+}\sec x_{2}^{+}x_{2}^{+}+\kappa\sec x_{1}^{+}(2x_{1}^{+}\sec x_{2}^{+}-x_{2}^{+}(x_{1}^{-}+\sec x_{2}^{+}))))]/\nonumber \\
 &  & \kappa\sec x_{1}^{+}x_{2}^{+}x_{1}^{-}\sec x_{2}^{-}(x_{2}^{-}-\sec x_{1}^{+})(\sec x_{1}^{-}x_{2}^{-}-\sec x_{1}^{+}x_{2}^{+})\nonumber \\
C & = & 2\sqrt{2}\sec{\gamma}_{1}\gamma_{2}\sec x_{1}^{-}x_{2}^{-}(\sec x_{1}^{+}x_{2}^{+}(x_{1}^{-}+\sec x_{2}^{+})-x_{1}^{+}\sec x_{2}^{+}(x_{2}^{-}+\sec x_{1}^{+}))/\nonumber \\
 &  & \kappa Gx_{1}^{-}\sec x_{2}^{-}(x_{2}^{-}-\sec x_{1}^{+})(\sec x_{1}^{-}x_{2}^{-}-\sec x_{1}^{+}x_{2}^{+})\nonumber \\
G & = & \frac{\gamma_{2}}{\sec{\gamma}_{2}}\frac{\sec x_{1}^{-}x_{2}^{-}\sec x_{2}^{+}}{x_{1}^{-}\sec x_{2}^{-}x_{2}^{+}}(\frac{x_{2}^{+}-x_{1}^{+}}{x_{2}^{-}-\sec x_{1}^{+}})\nonumber \\
H & = & \frac{\sec{\gamma}_{1}\sec x_{1}^{-}x_{2}^{-}\sec x_{2}^{+}}{\sec{\gamma}_{2}x_{1}^{-}\sec x_{2}^{-}x_{2}^{+}\sec x_{1}^{+}}(\frac{\sec x_{1}^{+}x_{2}^{+}-x_{2}^{-}x_{1}^{+}}{x_{2}^{-}-\sec x_{1}^{+}})\nonumber \\
K & = & \frac{\gamma_{2}\sec x_{1}^{-}x_{2}^{-}}{\gamma_{1}x_{1}^{-}\sec x_{2}^{-}x_{2}^{+}}(\frac{x_{1}^{+}\sec x_{2}^{+}-x_{1}^{-}x_{2}^{+}}{x_{2}^{-}-\sec x_{1}^{+}})\nonumber \\
L & = & \frac{\sec{\gamma}_{1}}{\gamma_{1}}\frac{\sec x_{1}^{-}x_{2}^{-}}{x_{1}^{-}\sec x_{2}^{-}\sec x_{1}^{+}x_{2}^{+}}(\frac{x_{2}^{-}x_{1}^{+}\sec x_{2}^{+}-x_{1}^{-}\sec x_{1}^{+}x_{2}^{+}}{x_{2}^{-}-\sec x_{1}^{+}})\nonumber \end{eqnarray}

The Yang-Baxter equation fails to hold for $g \neq \check g$, as already observed in the
 one-loop result of \cite{Gadde:2010zi}.

\subsubsection*{One-loop limit}

At one-loop, going back to the momentum representation,
the S-matrix simplifies to
\begin{eqnarray}
A & = & E\,=-\frac{1+e^{ip_{1}+ip_{2}}-2\kappa e^{ip_{2}}}{1+e^{ip_{1}+ip_{2}}-2\kappa e^{ip_{1}}}\\
B & = & D\,=-1 \nonumber\\
C & = & F\,=0 \nonumber\\
G & = & L\,=-\frac{\kappa(e^{ip_{1}}-e^{ip_{2}})}{1+e^{ip_{1}+ip_{2}}-2\kappa e^{ip_{1}}} \nonumber\\
H & = & K\,=-\frac{1+e^{ip_{1}+ip_{2}}-\kappa(e^{ip_{1}}+e^{ip_{2}})}{1+e^{ip_{1}+ip_{2}}-2\kappa e^{ip_{1}}}\nonumber
\end{eqnarray}
The S-matrix  $\check S$ for $\bar{Q}Q$ scattering  is given by sending
$\kappa\to\frac{1}{\kappa}$ in the above expressions. 

The bosonic and fermionic impurities do not mix at one-loop. The $Q$ $\bar Q$ S-matrix
 agrees  with the explicit perturbative calculation of \cite{Gadde:2010zi}.
The fermion S-matrix has also been successfully checked against  one-loop perturbation theory \cite{fullH}.

\subsubsection*{All-loops at $\kappa=0$}

For $\kappa =0$, the all-loop $S$ matrix at $\kappa=0$ in the $Q\bar{Q}$ channel
 is rather trivial,
\begin{eqnarray}
A & = & E\,=-1\\ \nonumber
B & = & D\,=-1\\ \nonumber
C & = & F\,=0\\ \nonumber
G & = & L\,=0\\ \nonumber
H & = & K\,=-1 \,.\nonumber
\end{eqnarray}
This is intuitively clear: the $Q$ and $\bar Q$ impurities are separated by adjoint fields in the ``checked'' vector multiplet,
which decouples in the limit $\kappa \to 0$.

On the other hand, in the $\bar{Q}Q$ scattering sector  the scattering retains
a a non-trivial dependence on the coupling (now the impurities are separated by the interacting fields of the ``unchecked'' vector multiplet),
\begin{center}
\begin{tabular}{ll}
$\check{A}=-e^{i(p_{2}-p_{1})}$ & $\check{D}=-1$\tabularnewline
$\check{B}=-e^{i(p_{2}-p_{1})}(\cos(p_{1}-p_{2})-i\frac{\sin(p_{1}-p_{2})}{\sqrt{1+2g^{2}}})\qquad$ & $\check{E}=-(\cos(p_{1}-p_{2})+i\frac{\sin(p_{1}-p_{2})}{\sqrt{1+2g^{2}}})$\tabularnewline
$\check{C}=-ie^{ip_{2}}\sqrt{2}g\frac{\sin(p_{1}-p_{2})}{\sqrt{1+2g^{2}}}$ & $\check{F}=-ie^{-ip_{1}}\sqrt{2}g\frac{\sin(p_{1}-p_{2})}{\sqrt{1+2g^{2}}}$\tabularnewline
$\check{G}=\frac{1}{2}(1-e^{i(p_{2}-p_{1})})$ & $ $$\check{L}=\frac{1}{2}(1-e^{i(p_{2}-p_{1})})$\tabularnewline
$\check{H}=-\frac{1}{2}(1+e^{i(p_{2}-p_{1})})$ & $\check{K}=-\frac{1}{2}(1+e^{i(p_{2}-p_{1})})$\,.\tabularnewline
\end{tabular}
\par\end{center}

The limit $\kappa \to 0$ is interesting because the $\mathbb{Z}_2$ quiver theory
reduces to ${\cal N}=2$ superconformal QCD (plus the decoupled ``checked'' vector multiplet).
We refer to \cite{Gadde:2009dj, Gadde:2010zi} for  detailed discussions. 
For $\kappa = 0$
the global symmetry $SU(2_{\hat I})$ combines with the second gauge group $SU(N_{\check c})$
and there is a symmetry enhancement to the flavor group
 $U(N_f = 2 N_c)$. 
 
 An important question is whether the flavor-singlet sector of the
  SCQQD spin-chain is integrable. We may now look forward to shed
  new light on this question using the above all-loop results.
  Unfortunately,
  flavor singlets are in particular $SU(2_{\hat I})$ singlets, and
   the methods of this paper     only allow us to consider scattering of 
   $SU(2_{\hat I})$ triplets. So our
   results have no direct bearing on the question of integrability of the ${\cal N}=2$ SQCD spin-chain.
With this caveat, we may nevertheless go ahead and check  whether the Yang-Baxter equation  holds at
$\kappa =0$ for $SU(2_{\hat I})$ triplets. It doesn't.\footnote{In \cite{Gadde:2010zi}, it was found that in the scalar sector, at one-loop,
 the YB equation holds as $\kappa \to 0$ both for $SU(2_{\hat I})$ triplets and $SU(2_{\hat I})$ singlets. Only the
 result for singlets is relevant to the integrability question.}

\section{\label{emergent}Emergent Magnons}

In \cite{Berenstein:2005jq}, following \cite{Berenstein:2005aa},
Berenstein et al$.$ reproduced the all-loop magnon dispersion relation in ${\cal N}=4$ SYM 
using a simple matrix quantum mechanics.
The matrix quantum mechanics is obtained by truncating to the lowest
modes of $SU(N_c)$ ${\cal N}=4$ SYM on $S^{3}$. The ground state
 is obtained by minimizing the potential energy, which
leads to a model of commuting hermitian matrices. The matrix eigenvalues
are localized on a five-sphere of radius\footnote{Our  normalization for the fields
are related to the normalization 
in \cite{Berenstein:2005jq} as $\phi_{here}=\phi_{there}/\sqrt{N_c}$.} 
 $\frac{1}{\sqrt 2}$,
which is naturally identified with the $S^{5}$ in the dual background. 
This gives a simple picture for emergent geometry. Each point in the emergent geometry corresponds
to an eigenvalue and is labelled by an $SU(N_c)$ index.
 In \cite{Berenstein:2005ek, Berenstein:2006yy}
 the same  exercise for  orbifolds of ${\cal N}=4$ SYM shows 
that the ground state of the matrix model  is localized on the orbifolded
$S^5$. 

\begin{figure}[h!]
\begin{center}
\includegraphics[scale=0.4]{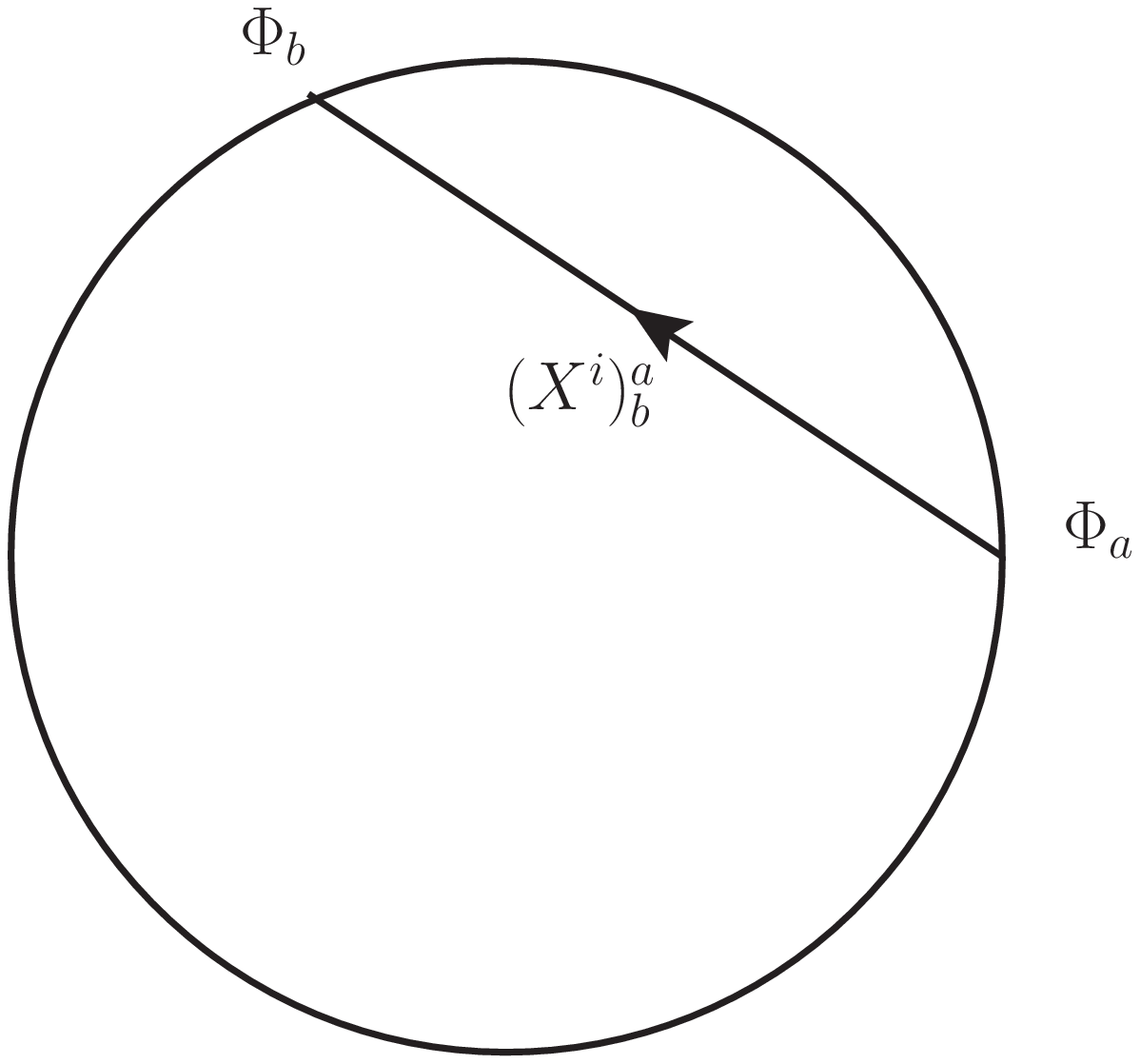}$\qquad\qquad$\includegraphics[scale=0.4]{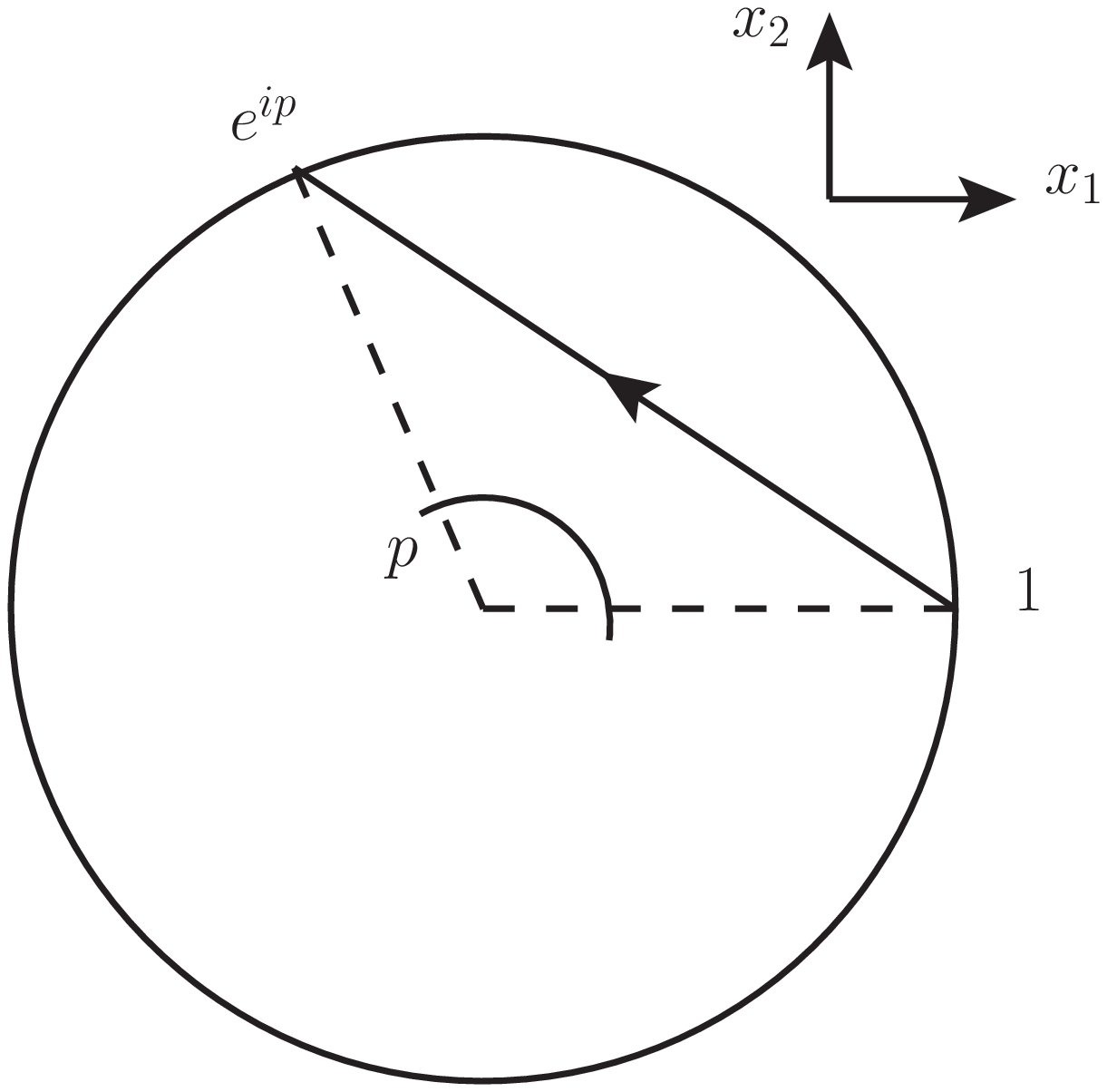}
\par\end{center}
\caption{\label{N4stringbits} The left figure shows the string bit corresponding to the off-diagonal excitation $(X^i)^a_b$. The right figure shows  the configuration where the wavefunction of a  magnon with momentum $p$ is peaked.}
\end{figure}

The excitations of the vacuum obtained by turning on off-diagonal modes
of the matrix model are interpreted as string bits. They are bilocal in the emergent geometry
 because they are labelled by two $SU(N_c)$ indices and are visualized
as string bits stretching
between two points (see figure \ref{N4stringbits}). An off-diagonal excitation of momentum
$p$  is peaked at the configuration where the corresponding
string bit subtends an angle $p$ at the center. The expectation value
of their energy precisely reproduces the exact magnon dispersion relation \cite{Berenstein:2005jq}.
A very similar picture for
the magnons was obtained in \cite{Hofman:2006xt}
on the dual string side. 
Moreover, the $x_{1}$ and $x_{2}$ components 
of the vector $\vec M$ associated with the magnon
were identified with the  central charges of the $SU(2|2)$ algebra \cite{Hofman:2006xt}
\[ M_1 = \frac{1}{2} (K + P) \, , \quad M_2 = \frac{1}{2 i} (K - P)\,. \label{Mcentral}
\]

\subsection{Emergent magnons for the $\mathbb{Z}_2$ quiver}

\begin{figure}[h!]
\begin{center}
\includegraphics[scale=0.4]{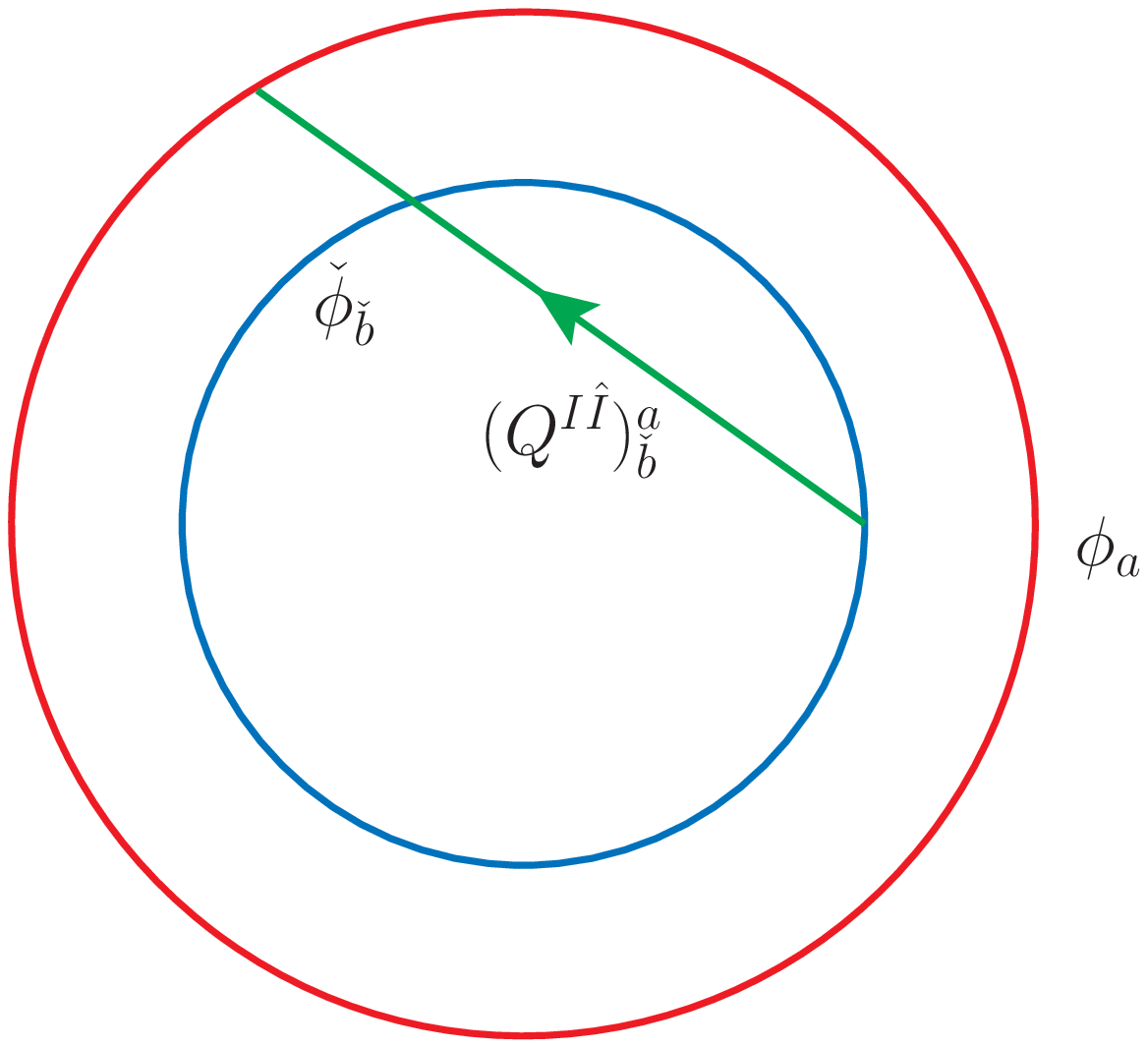}$\qquad\qquad$\includegraphics[scale=0.4]{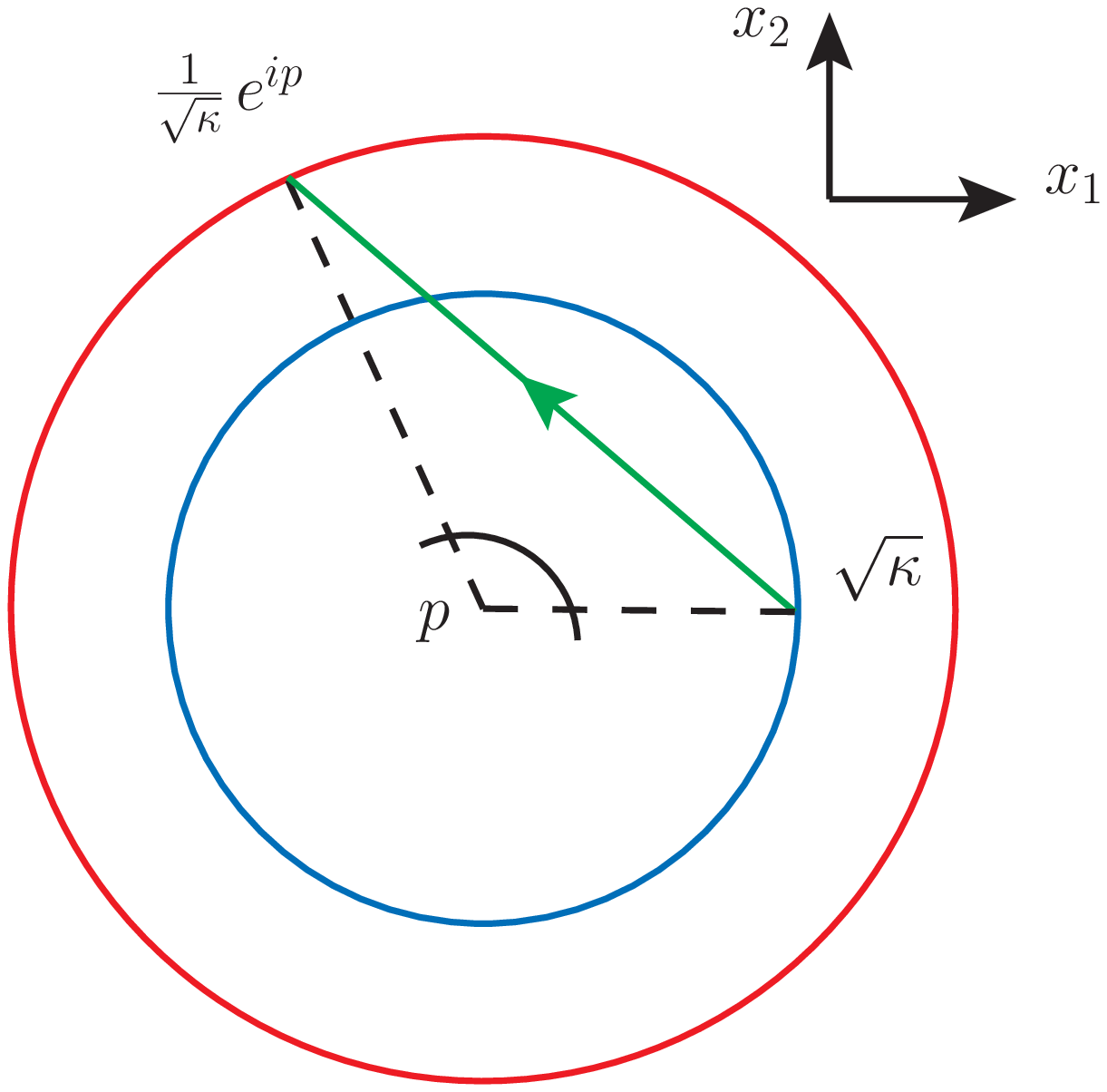}
\par\end{center}
\caption{\label{interpolating}The figure on the left shows the string bit in the $\mathbb{Z}_2$ quiver theory.
 On the right, the wavefunction of the bifundamental magnon $Q^{I\hat I}$ with momentum $p$.}
\end{figure}

Following \cite{Berenstein:2005jq}, we truncate the  $\mathbb{Z}_2$ quiver
theory to its lowest bosonic modes on $S^{3}$, which gives us the matrix quantum mechanis
\begin{eqnarray}
S & = & N_c \int dt\, \mbox{Tr}\:\frac{1}{2}\left((D_{t}\phi)^{2}+(D_{t}\check{\phi})^{2}+(D_{t}Q^{I\hat{I}})^{2}-\phi^{2}-\check{\phi}^{2}-(Q^{I\hat{I}})^{2}\right)\\
 & - & g^2 \left([\phi,\bar{\phi}]^{2}+\sqrt{2}Q^{I\hat{I}}\bar{Q}_{I\hat{I}}(\phi\bar{\phi}+\bar{\phi}\phi)+Q^{I\hat{I}}\bar{Q}_{J\hat{I}}Q^{J\hat{J}}\bar{Q}_{I\hat{J}}-\frac{1}{2}Q^{I\hat{I}}\bar{Q}_{I\hat{I}}Q^{J\hat{J}}\bar{Q}_{J\hat{J}}\right) \nonumber\\
 & - & {\check g}^2 \left([\check{\phi},\bar{\check{\phi}}]^{2}+\sqrt{2}\bar{Q}_{I\hat{I}}Q^{I\hat{I}}(\check{\phi}\bar{\check{\phi}}+\bar{\check{\phi}}\check{\phi})+\bar{Q}_{J\hat{I}}Q^{I\hat{I}}\bar{Q}_{I\hat{J}}Q^{J\hat{J}}-\frac{1}{2}\bar{Q}_{I\hat{I}}Q^{I\hat{I}}\bar{Q}_{J\hat{J}}Q^{J\hat{J}}\right) \nonumber\\
 & + & \sqrt{g\check g }\left(4Q^{I\hat{I}}\check{\phi}\bar{Q}_{I\hat{I}}\bar{\phi}+h.c.\right)+\frac{1}{N_{c}}
 (\mbox{double}-\mbox{ trace}).\nonumber
 \end{eqnarray}
The mass terms arise due to the conformal couplings of the scalars
to curvature of $S^{3}$. The eigenvalue distribution of the ground state
is same as that of the $\mathbb{Z}_{2}$ orbifold of ${\cal N}=4$
SYM. We now excite the off-diagonal mode $(Q^{I\hat{I}})_{\check{b}}^{a}$.
The linearized theory describing this excitation is the harmonic oscillator,
\begin{eqnarray*}
H & = & \frac{1}{2}(\Pi_{I\hat{I}})_{\check{b}}^{a}(\Pi^{I\hat{I}})_{a}^{\check{b}}+\frac{1}{2}\omega_{a\check{b}}(Q^{I\hat{I}})_{\check{b}}^{a}(\bar{Q}_{I\hat{I}})_{a}^{\check{b}}\\
\omega_{a\check{b}} & = & 1+4|g \phi_{a}-\check{g} \check{\phi}_{\check{b}}|^{2}.
\end{eqnarray*}
Note the difference in the frequency compared to the ${\cal N}=4$ case, where
$\omega_{ab}=1+4g^2|\phi_{a}-\phi_{b}|^{2}$. This
motivates the effective picture of figure \ref{interpolating}. 

The circle spanned by the eigenvalues of $\Phi$ has split into two
circles, one spanned by the eigenvalues of $\phi$ and the other by
eigenvalues of $\check{\phi}$. The radii of the two circles are taken
to be $\frac{1}{\sqrt{\kappa}}\frac{G}{\sqrt{2}}$ and $\sqrt{\kappa}\frac{G}{\sqrt{2}}$
respectively, by normalizing the tension of the string bit to unity.
The string bit corresponding to a bifundamental excitation stretches
from one circle to the other.  A magnon of momentum $p$ again
localizes on the configuration where the string bit subtends an angle
$p$ at the center. Using (\ref{Mcentral})
we learn\[
{P}=x_{1}-ix_{2}=\frac{G}{\sqrt{2}}(e^{-ip}\frac{1}{\sqrt{\kappa}}-\sqrt{\kappa})={ K}^{*}\, ,\]
so the energy of the magnon is \[
\Delta-|r|=\sqrt{1+8G^{2}\left(\sin^{2}\frac{p}{2}+\frac{1}{4}(\sqrt{\kappa}-\frac{1}{\sqrt{\kappa}})^{2}\right)} .\]
The central charges 
 agree precisely with the from obtained earlier from the 
algebraic method.\footnote{Of course, as before, there is no guarantee
that the couplings do not get renormalized. This caveat is all the more obvious
in this approach, since integrating out  massive modes would generically
lead to such a renormalization.}

\begin{figure}
\begin{center}
\includegraphics[scale=0.4]{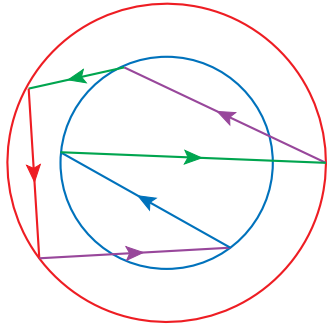}
\par\end{center}
\caption{\label{generic} A  state of the spin chain with six magnons.}
\end{figure}

 It is
clear that the adjoint excitations $\lambda$ and $D$ ($\check{\lambda}$ and $\check D$) are string bits that stretch between two points
of $\phi$ circle ($\check{\phi}$ circle). Their dispersion relation
coincides with the ${\cal N}=4$ SYM dispersion relation, as clear from the
picture. A generic state of the spin chain is shown in figure \ref{generic}.

At strong 't Hooft coupling, 
Hofman and Maldacena \cite{Hofman:2006xt} obtained the dual description of an ${\cal N}=4$ magnon  as a  semiclassical strings rotating on the $S^2 \subset S^5$. In LLM coordinates this ``giant magnon''  has precisely
the shape of figure \ref{N4stringbits}. The energy of the string was matched with the strong coupling
limit of the exact magnon dispersion relation. (See also \cite{Arutyunov:2006ak} for a sigma-model
derivation of the $SU(2|2)$ central charges.)
The ${\mathbb Z}_2$ quiver theory is dual to the $AdS_5\times S^5/{\mathbb Z}_2$ background. The ratio
of the gauge couplings is related the period of the NSNS B-field through the collapsed
two-cycle. It must be possible to reproduce the effective picture of figure \ref{interpolating} and the associated
dispersion relation by studying the giant magnon solution in this background. This problem is under investigation
 \cite{MagnonBfield}. 

\subsection{Bound states}

In addition to the elementary magnons with real momenta, the spectrum of the theory also contains bound states
at some special complex values of the momenta. A two-magnon bound state occurs
at the pole of the two-body S-matrix, \[
S(p_{1},p_{2})=\infty\qquad\mbox{with }\quad p_{1}=\frac{P}{2}-iq,\quad p_{2}=\frac{P}{2}+iq,\quad q>0\,.\]
 Since
$S(p_{2},p_{1})=1/S(p_{1},p_{2})\to0$, the asymptotic wavefunction
becomes\[
e^{iP\frac{x_{1}+x_{2}}{2}-q|x_{2}-x_{1}|}.\]
A bound state has smaller energy than any state in the two particle
continuum with the same total mometum $P$. The exact dispersion relation
of the bound states in ${\cal N}=4$ SYM was found in \cite{Dorey:2006dq}
and their S-matrix in \cite{Roiban:2006gs}. The two-body S-matrix
in the present case allows us to determine the bound state dispersion
relation. Finding their S-matrix, however,  would requires the four-body magnon S-matrix,
which we cannot determine in the absence of integrability.

Let us first analyze the bound state of $Q^{+}$ (on the left of the chain) and
$\bar{Q}^{+}$ (on the right). Their scattering matrix given
in equ.(\ref{eq:solution}), \[
A(p_{1},p_{2})=S_{12}^{0}\frac{\sec x_{2}^{-}x_{1}^{-}}{x_{2}^{-}\sec x_{1}^{-}}(\frac{\sec x_{1}^{+}-x_{2}^{-}}{x_{1}^{-}-\sec x_{2}^{+}}) \, , \]
where $S_{12}^{0}$ is the overall dressing factor which is not determined
by symmetries. Clearly there is a pole is at $x_{1}^{-}=\tilde{x}_{2}^{+}$.
We assume that this pole is not cancelled by a zero of the dressing
factor. Following \cite{Dorey:2007an}, we define the bound state
rapidity variables as \[
X^{+}\equiv x_{1}^{+},\qquad X^{-}\equiv\sec x_{2}^{-} \, .\]
Remarkably, at the pole they obey the relations\begin{eqnarray*}
\frac{X^{+}}{X^{-}} & = & e^{iP}\\
X^{+}-X^{-}+\frac{G^{2}}{2}(\frac{1}{X^{+}}-\frac{1}{X^{-}}) & = & 2i\sqrt{\kappa}.\end{eqnarray*}
The bound state dispersion relation can also be expressed completely
in terms of $X^{\pm}$,
\begin{eqnarray}
C_{Q\bar{Q}} & = & C_{1}+\tilde{C}_{2}=1+i\frac{G^{2}}{2\sqrt{\kappa}}(\frac{1}{X^{+}}-\frac{1}{X^{-}})
\nonumber \\
 & = & \frac{1}{2}\sqrt{4+8g^{2}\sin^{2}\frac{p}{2}}.
 \end{eqnarray}
This dispersion is exactly the same as the one of the two-magnon bound
states in ${\cal N}=4$ SYM. Thus the $Q\bar{Q}$ bound state can
be elegantly represented as a string bit of {}``weight two'' stretching
between two points of the outer circle. The analogous exercise for the $\bar{Q}Q$
bound state gives the dispersion relation\[
C_{\bar{Q}Q}=\frac{1}{2}\sqrt{4+8\check{g}^{2}\sin^{2}\frac{p}{2}}\, .\]
This bound state is represented as a weight-two string bit stretching
between two points of the inner circle. 

As we vary the momentum $P$
of the bound state the pole $iq$ moves on the positive imaginary
axis. For certain values of $P$ where $q$ approaches zero, the bound
state is only marginally stable. This phenomenon does not occur in
${\cal N}=4$ SYM, the bound states of ${\cal N}=4$ are stable for
all values of $P$ but this is not the case for the $\mathbb{Z}_2$ quiver theory. The 
marginal stability condition $q=0$ gives respectively for the $Q\bar{Q}$ and $\bar{Q}Q$ bound
states \[
\kappa=\cos\frac{P}{2}\qquad\mbox{and}\qquad\frac{1}{\kappa}=\cos\frac{P}{2}\]
In the latter case, there is no solution which means
that $\bar{Q}Q$ bound state is stable for all values of the momenta. On the other hand,
the $Q\bar{Q}$ bound state on the other hand can decay at $P=2\arccos\kappa$.
These conclusions exactly match with results obtained at one loop in \cite{Gadde:2010zi}.

\begin{figure}
\begin{center}
\includegraphics[scale=0.35]{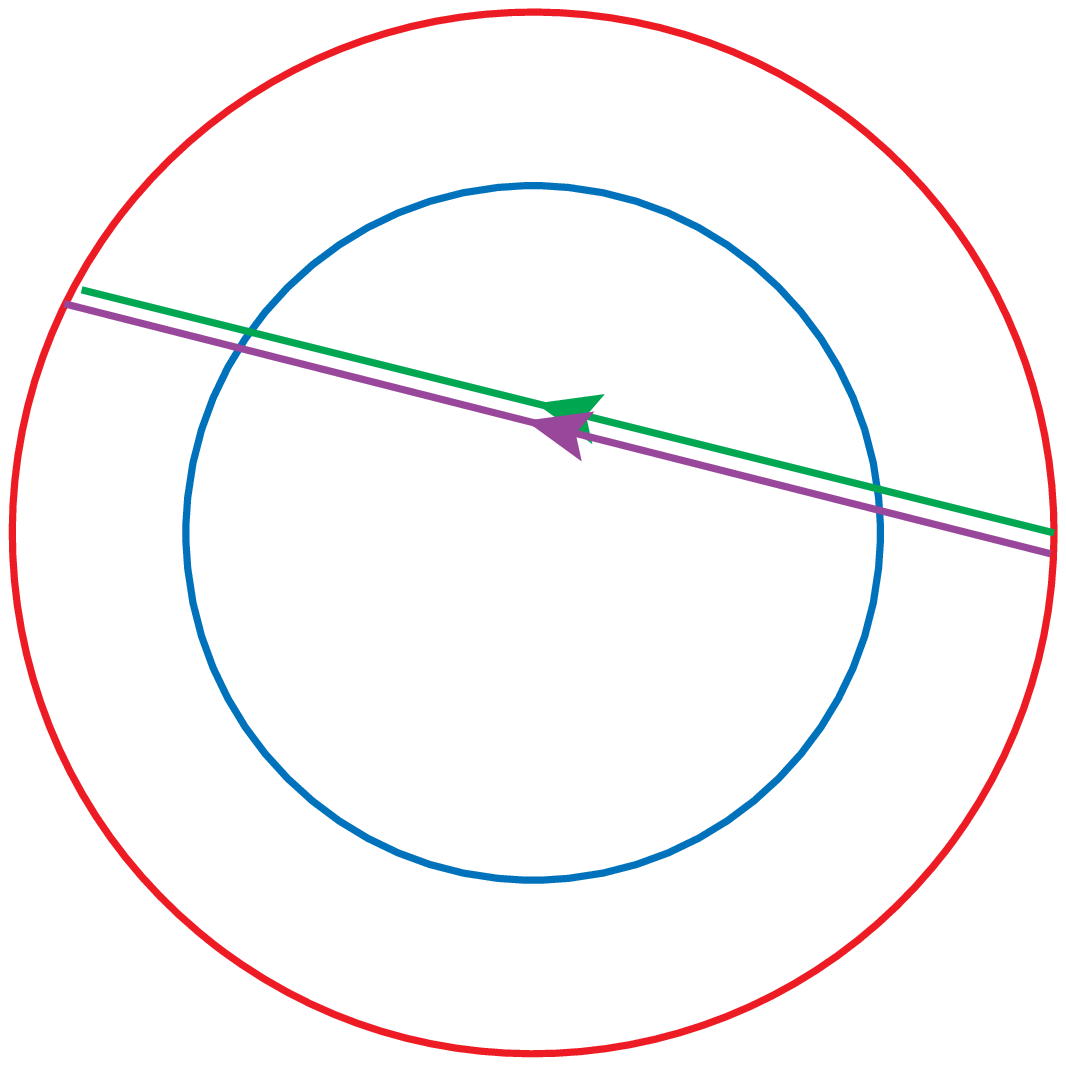}$\qquad$\includegraphics[scale=0.35]{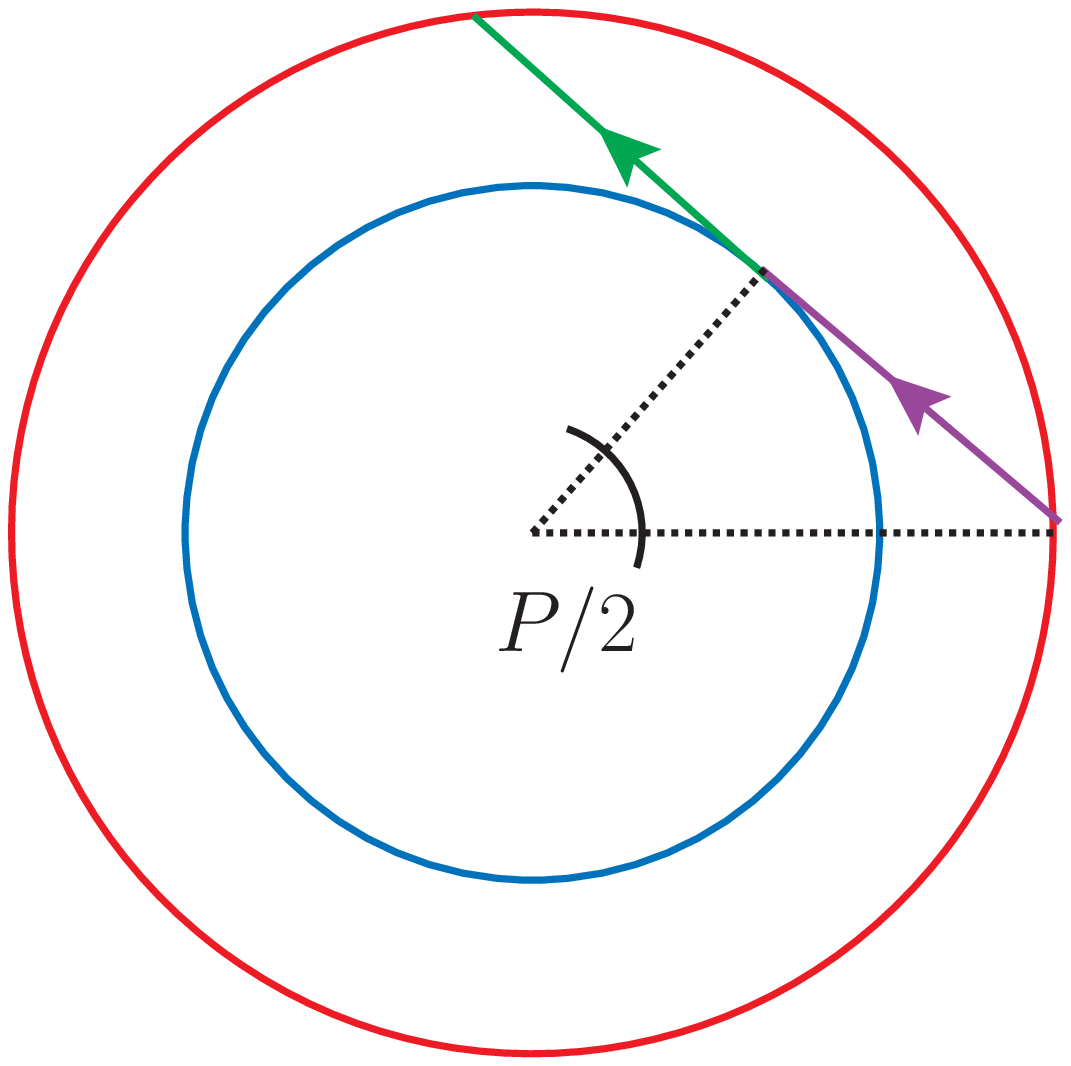}$\qquad$\includegraphics[scale=0.35]{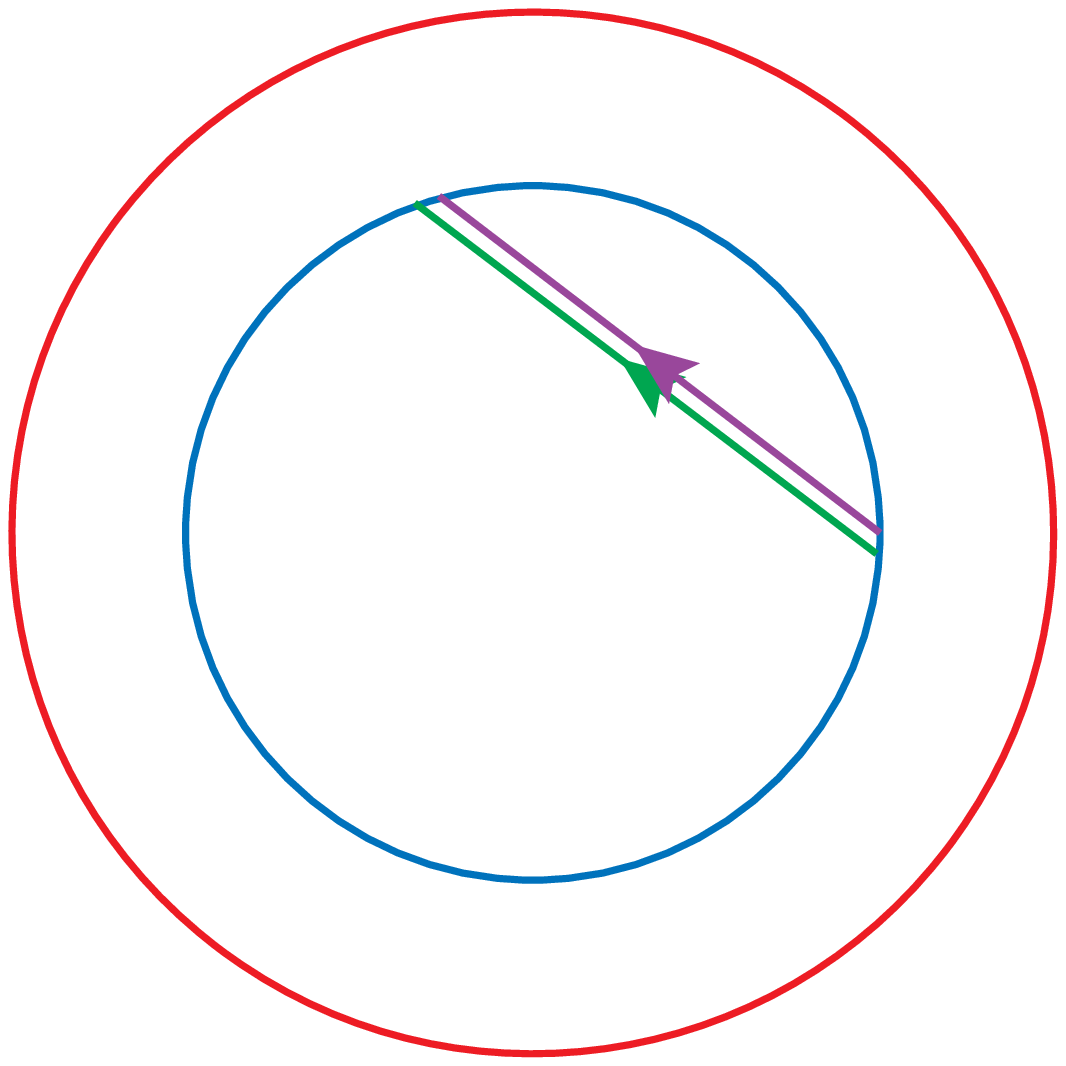}
\par\end{center}
\caption{\label{bound}The figure on the left represents a $Q\bar Q$ bound state at generic momenta. In the middle is the marginally stable $Q \bar Q$ bound state. From the figure one can easily see that $P=2\arccos \kappa$ since 
the ratio of the radii of the two circles is $\kappa$. On the right is a   $\bar Q Q$ bound state, which is stable for all values of momenta.}
\end{figure}

Geometrically, there is simple way of understanding the boud state
decay, see figure \ref{bound}. As the bound state string bit stretching in the outer circle
(which means it is a $Q\bar{Q}$ bound state) touches the inner circle,
its energy becomes manifestly equal to the sum
of the energies of the constituents. Vanishing of the binding energy
allows the $Q\bar{Q}$ state to decay. Simple trigonometry reveals the
threshold momentum $P=2\arccos\kappa$ at this point. From this picture it
is also immediate to see that the $\bar{Q}Q$ bound state is stable for all values
of the momenta.

As we move around in the parameter space of the quiver gauge theory, at certain codimension one ``walls'', the bound states of the elementary magnons decay. It would be interesting to understand  bound state decay as a wall-crossing phenomenon
in the dual sigma model.

\section{Generalization to $\mathbb{Z}_{k}$ orbifolds}
\label{Z_k}
\begin{figure}
\begin{center}
\includegraphics[scale=0.4]{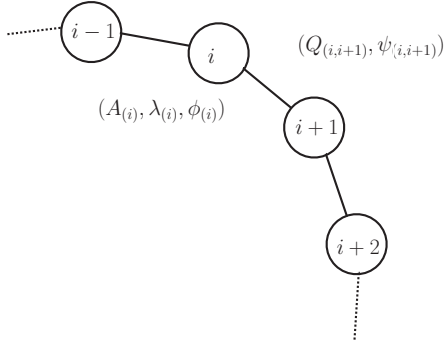}
\par\end{center}
\caption{\label{quiver}The quiver diagram for ${\cal N}=2$ ${\mathbb Z}_2$ orbifold of ${\cal N}=4$ SYM. It is a circular necklace with $k$ nodes, four of which are shown. A vector multiplet $(A,\lambda,\phi)$ is associated to each node and a hypermultiplet $(Q^I,\psi)$ is associated to each edge.}
\end{figure}

The analysis presented for the $\mathbb{Z}_{2}$ quiver can be extended
to a general ADE ${\cal N}=2$ orbifold of ${\cal N}=4$ SYM. In this section we  indicate
the generalization for the (marginally deformed)
 $\mathbb{Z}_{k}$ orbifolds. The quiver gauge
theory describing such an orbifold is shown in figure \ref{quiver}.

The superpotential
at a generic point in the parameter space is\[
W=\frac{1}{\sqrt{2}}\sum_{i}g_{(i)}\left(\mbox{Tr}Q_{(i-1,i)}^{I}\phi_{(i)}\bar{Q}_{I(i,i-1)}+\mbox{Tr}\bar{Q}_{I(i+1,i)}\phi_{(i)}Q_{(i,i+1)}^{I}\right).\]
We  impose the periodicity condition $i+k\sim i$ on the indices.

To compute the $SU(2|2)$ central charges for the representation of the $Q_{(i,i+1)}^{I}$
magnon  we evaluate the anticommutator of two supersymmetries,\[
\{Q_{I}^{\dot{\alpha}},Q_{J}^{\dot{\beta}}\}Q_{(i,i+1)}^{K}=\epsilon^{\dot{\alpha}\dot{\beta}}\epsilon_{IJ}(\frac{g_{(i)}}{\sqrt{2}}\phi_{(i)}Q_{(i,i+1)}^{K}-\frac{g_{(i+1)}}{\sqrt{2}}Q_{(i,i+1)}^{K}\phi_{(i+1)})\]
which, on the spin chain, leads to\begin{eqnarray*}
\{Q_{I}^{\dot{\alpha}},Q_{J}^{\dot{\beta}}\}|Q_{(i,i+1)}^{K}\rangle & = & \epsilon^{\dot{\alpha}\dot{\beta}}\epsilon_{IJ}\frac{1}{\sqrt{2}}(g_{(i)}e^{-ip}-g_{(i+1)})|Q_{(i,i+1)}^{K}\phi^{+}\rangle\\
\Rightarrow\gen P & = & \frac{1}{\sqrt{2}}(g_{(i)}e^{-ip}-g_{(i+1)})=\gen K^{*}.\end{eqnarray*}
Interchanging $g_{(i)}\leftrightarrow g_{(i+1)}$ gives us the central
charges of the $\bar{Q}_{(i+1,i)}$ representation. In both cases
we get the dispersion relation\[
\Delta-|r|=2\gen C=\sqrt{1+8G_{(i,i+1)}^{2}\left(\sin^{2}\frac{p}{2}+\frac{1}{4}(\sqrt{\kappa_{(i,i+1)}}-\frac{1}{\sqrt{\kappa_{(i,i+1)}}})^{2}\right)}\,.\]
Here we have defined \[
G_{(i,i+1)}=\sqrt{g_{(i)}g_{(i+1)}}\qquad\mbox{and}\qquad\kappa_{(i,i+1)}=\frac{g_{(i+1)}}{g_{(i)}}.\]
The dispersion relation of the adjoint magnons $\lambda_{(i)}$ and
$D_{(i)}$ works the same way as ${\cal N}=4$ and is equal to\[
\Delta-|r|=2\gen C=\sqrt{1+8g_{(i)}^{2}\sin^{2}\frac{p}{2}}.\]

\begin{figure}
\begin{center}
\includegraphics[scale=0.4]{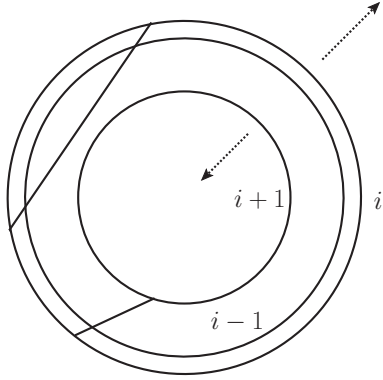}
\par\end{center}
\caption{\label{generalpicture}The emergent picture describing
$\mathbb{Z}_{k}$ orbifold. Only the circles corresponding to $i-1,i,i+1$
gauge node are shown. We have also shown two magnons, one in the adjoint
of $SU(N)_{i}$ and the other in the bifundamental of $SU(N)_{i}\times SU(N)_{i+1}$.}
\end{figure}

The picture presented in section \ref{emergent} also generalizes
to $\mathbb{Z}_{k}$ orbifolds, see figure \ref{generalpicture}. It consists of $k$ concentric circles
which are labelled by $i$, corresponding to the gauge group $SU(N_c)_{i}$.
The radius of $i$-th circle is $\frac{g_{(i)}}{\sqrt{2}}$. The magnons
in the adjoint of the $i$-th node are represented by string bits that
stretch between the $i$-th circle, while the $SU(N)_{i}\times SU(N)_{i+1}$
bifundamental magnons correspond to string bits
stretching from $i$-th to $i+1$-th circle.
The dispersion relations of both adjoint and bifundamental magnons
is summarized by the simple formula\[
\Delta-|r|=\sqrt{1+4\ell^{2}}\]
where $\ell$ is the length of the corresponding string bit.  The two-body S-matrix
 is also fixed by the centrally extended $SU(2|2)$ symmetry, and can be
obtained by straightforward extension of our analysis of the $\mathbb{Z}_2$ case.

\section*{Acknowledgements}

It is a pleasure to thank David Berenstein, Nikolay Bobev, Pedro Liendo, Elli Pomoni, Pedro Vieira and Wenbin Yan  for useful discussions.
This work was supported in part by DOE grant DEFG-0292-ER40697 and by NSF grant PHY-0653351-001.
Any opinions, findings, and conclusions or recommendations expressed in this material are those of the authors and do not necessarily reflect the views of the National Science Foundation.

\appendix

\section{Algebraic constraints on the central charges \label{sec:Beisertmethod}}

\subsection{${\cal N}=4$ super Yang-Mills}

Let us review the logic used in
 \cite{Beisert:2005tm} to constrain the central elements $\gen P$ and $\gen K$.
 The action of 
$\gen P$ on a state with $K$ $\mathcal{X}$-excitations with momenta
$p_{1},\ldots p_{K}$ is 
\begin{equation}
\gen{P|}\mathcal{X}_{1}\mathcal{X}_{2}\ldots\mathcal{X}_{K}\rangle=\sum_{k=1}^{K}a_{k}b_{k}\prod_{l=k+1}^{K}e^{-ip_{l}}|\mathcal{X}_{1}\mathcal{X}_{2}\ldots\mathcal{X}_{K}\Phi^{+}\rangle\label{eq:totalP}\end{equation}
On a physical state like the one above, the central charge must vanish.
Since in the ${\cal N}=4$ case all the
$\mathcal{X}$-excitations belong to the same (fundamental) representation of $SU(2|2)$,
the central
charge only depends upon the momentum and not on the type of excitation,
and the only possibility is  for the sum in (\ref{eq:totalP}) to telescope
to zero on physical states, \[
a_{i}b_{i}=\alpha(e^{-ip_{i}}-1)\equiv P\]
with $\alpha$ being an undetermined constant. Here we use the fact
that the total momentum of a physical state is zero. A similar exercise
for $\gen K$ gives \[
c_{i}d_{i}=\beta(e^{ip_{i}}-1)\equiv K \, .\]
On a single-particle state,\begin{equation}
\gen{P|}{\cal X}\rangle=\alpha(e^{ip}-1)|{\cal X}\Phi^{+}\rangle,\qquad\gen{K|}{\cal X}\rangle=\beta(e^{-ip}-1)|{\cal X}\Phi^{-}\rangle \, .\label{eq:hermitian1}\end{equation}
The hermiticity condition translates into $\alpha=\beta^{*}$. 
Finally \[
C=\frac{1}{2}\sqrt{1+4PK} = \frac{1}{2}\sqrt{1+16\alpha\beta\sin^{2}\frac{p}{2}}\,.\]
Comparing with the one loop dispersion relation one finds $\alpha\beta=\frac{g^{2}}{2}+O(g^{4}) \equiv \frac{{\bf g}^2}{2}$.

\subsection{$\mathbb{Z}_2$ quiver}

A physical state is constructed by having  alternating $Q$ and $\bar{Q}$ type impurities
 on a periodic spin chain. The central charge should vanish
on such a state. To determine the central
charges $\gen P$ and $\gen K$ as functions of magnon mometum, we
follow same steps as before.  The action of $\gen P$ and $\gen K$
is \begin{eqnarray*}
&&\gen P| Q_1 {\bar Q}_2 \ldots Q_{K-1} {\bar Q}_K \rangle \\
&& =\, (a_{1}b_{1}(e^{-ip_{2}}\ldots e^{-ip_{K}})+\sec a_{2}\sec b_{2}(e^{-ip_{3}}\ldots e^{-ip_{K}})+\ldots+\sec a_{K}\sec b_{K}) | Q_1 {\bar Q}_2 \ldots Q_{K-1} {\bar Q}_K \phi^+\rangle\\
&&\gen K | Q_1 {\bar Q}_2 \ldots Q_{K-1} {\bar Q}_K \rangle \\
&& =\, (c_{1}d_{1}(e^{ip_{2}}\ldots e^{ip_{K}})+\sec c_{2}\sec d_{2}(e^{ip_{3}}\ldots e^{ip_{K}})+\ldots+\sec c_{K}\sec d_{K}) | Q_1 {\bar Q}_2 \ldots Q_{K-1} {\bar Q}_K \phi^-\rangle.\end{eqnarray*}
As before, let us define $P_i \equiv a_i b_i, K_i \equiv c_i d_i$ and ${\sec P}_i \equiv {\sec a}_i {\sec b}_i, \sec K_i \equiv \sec c_i \sec d_i$.
Now we impose
\begin{enumerate}
\item Physical state condition: \\
$\gen P$ and $\gen K$ should vanish when the total momentum of the
state is zero. 
\item BPS condition: \\ 
A BPS state of the interpolating theory is obtained from a BPS state of the orbifold by the substitution  (in the one-loop approximation)
$\check \phi \to \kappa \check \phi$, $\kappa \equiv {\check g} /g$ (see the last paragraph of appendix  B in \cite{Gadde:2009dj}).
At higher orders we may have a renormalized substitution $\check \phi \to \kappa' \check \phi$,
 $k'  \equiv {\bf \check g}/{\bf g}$ with ${\bf g}(g, \check g)$ and ${\bf\check g}(g, \check g)$
 renormalized couplings.
This means $Q(\bar Q)$ moving with momentum $i\ln\kappa'\,\,(-i\ln\kappa')$ is chiral 
and we  expect that $P_{i}K_{i}\,\,(\sec P_{i}\sec K_{i})$ should vanish on that state.
\item Hermiticity: \\
$K=P^*$ and ${\tilde K}={\tilde P}^*$.
\end{enumerate}
From these condition it follows that
\begin{eqnarray*}
P & = & \alpha(e^{-ip}\frac{1}{\sqrt{\kappa'}}-\sqrt{\kappa'}),\qquad\qquad K=\alpha^*(e^{ip}\frac{1}{\sqrt{\kappa'}}-\sqrt{\kappa'}),\\
\sec P & = & {\alpha}(e^{-ip}\sqrt{\kappa'}-\frac{1}{\sqrt{\kappa'}}),\qquad\qquad\sec K= {\alpha}^*(e^{ip}\sqrt{\kappa'}-\frac{1}{\sqrt{\kappa'}}).
\end{eqnarray*}
($\{P,K\}\leftrightarrow \{\sec P,\sec K\}$ is of course also a solution since the conditions
above make no intrinsic distinction between the $Q$ and $\bar Q$ impurities.)
We then have
\begin{eqnarray}
C=\frac{1}{2}\sqrt{1+4PK} &=& \frac{1}{2}\sqrt{1+16|\alpha|^2\Big(\sin^{2}\frac{p}{2}+\frac{1}{4}(\sqrt{\kappa'}-\frac{1}{\sqrt{\kappa'}})\Big)^2}\,\\
\sec C=\frac{1}{2}\sqrt{1+4\sec P \sec K} &=& \frac{1}{2}\sqrt{1+16|\alpha|^2\Big(\sin^{2}\frac{p}{2}+\frac{1}{4}(\sqrt{\kappa'}-\frac{1}{\sqrt{\kappa'}})\Big)^2}.
\end{eqnarray}
Comparing with the one-loop dispersion relation \cite{Gadde:2010zi}
one finds $|\alpha|^2  \equiv \frac{ {\bf g \,\check g }}{2}=\frac{g \check g}{2}+\ldots$.  All in all, 
\[
C = \sec C =  \sqrt{1+2 ({\bf g}-{{\bf \check g} })^2+8 {\bf g \check g} \sin^{2}\frac{p}{2}}\,.
\]

\section{\label{derive}Solving for the S-matrix}

\subsubsection*{$SU(2|1)$ subsector: Determining $A,\, K,\, G,\: H,\: L$}

We first consider the $SU(2_{\dot{\alpha}}|1_{I})$ subsector,
which is closed under scattering.
 Consider the scattering of two bosonic
magnons $Q^{+}$ and $\bar{Q}^{+}$. Requiring invariance under the supercharge  ${\cal Q}_{+}^{\dot{\alpha}}$
we find
\begin{eqnarray*}
\gen Q_{\,+}^{\dot{\alpha}}S_{12}|Q_{1}^{+}\bar{Q}_{2}^{+}\rangle & = & \gen Q_{+}^{\dot{\alpha}}A_{12}|Q_{2}^{+}\bar{Q}_{1}^{+}\rangle\\
 & = & A_{12}a_{2}|\psi_{2}^{\dot{\alpha}}\bar{Q}_{1}^{+}\rangle+A_{12}\sec a_{1}|Q_{2}^{+}\tilde{\psi}_{1}^{\dot{\alpha}}\rangle\\
S_{12}\gen Q_{\:+}^{\dot{\alpha}}|Q_{1}^{+}\bar{Q}_{2}^{+}\rangle & = & S_{12}(a_{1}|\psi_{1}^{\dot{\alpha}}\bar{Q}_{2}^{+}\rangle+\sec a_{2}|Q_{1}^{+}\tilde{\psi}_{2}^{\dot{\alpha}}\rangle)\\
 & = & (a_{1}K_{12}+\sec a_{2}G_{12})|\psi_{2}^{\dot{\alpha}}\bar{Q}_{1}^{+}\rangle+(a_{1}L_{12}+\sec a_{2}H_{12})|Q_{2}^{+}\tilde{\psi}_{1}^{\dot{\alpha}}\rangle)\\{}
[\gen Q_{+}^{\dot{\alpha}},S]=0 & \Rightarrow\\
A_{12} & = & \frac{a_{1}}{a_{2}}K_{12}+\frac{\sec a_{2}}{a_{2}}G_{12}\\
A_{12} & = & \frac{a_{1}}{\sec a_{1}}L_{12}+\frac{\sec a_{2}}{\sec a_{1}}H_{12}.\end{eqnarray*}
More constraints are obtained by imposing invariance under conformal
supersymmetries $\gen S$. In this subsector 
 it is sufficient to focus on $\gen S_{\dot{\alpha}}^{-}$,
\begin{eqnarray*}
\gen S_{\:\dot{\alpha}}^{-}S_{12}|Q_{1}^{+}\bar{Q}_{2}^{+}\rangle & = & A_{12}(-c_{2}\epsilon_{\dot{\alpha}\dot{\beta}}|\psi_{2}^{\dot{\beta}}\check{\phi}^{-}\bar{Q}_{1}^{+}\rangle-\sec c_{1}\epsilon_{\dot{\alpha}\dot{\beta}}|Q_{2}^{+}\tilde{\psi}_{1}^{\dot{\beta}}\phi^{-}\rangle)\\
 & = & A_{12}(-c_{2}\epsilon_{\dot{\alpha}\dot{\beta}}\frac{x_{2}^{-}}{x_{2}^{+}}|\phi^{-}\psi_{2}^{\dot{\beta}}\bar{Q}_{1}^{+}\rangle-\sec c_{1}\epsilon_{\dot{\alpha}\dot{\beta}}\frac{x_{2}^{-}\sec x_{1}^{-}}{x_{2}^{+}\sec x_{1}^{+}}|\phi^{-}Q_{2}^{+}\tilde{\psi}_{1}^{\dot{\beta}}\rangle)\\
S_{12}\gen S_{\:\dot{\alpha}}^{+}|Q_{1}^{+}\bar{Q}_{2}^{+}\rangle & = & S_{12}(-c_{1}\epsilon_{\dot{\alpha}\dot{\beta}}\frac{x_{1}^{-}}{x_{1}^{+}}|\phi^{-}\psi_{1}^{\dot{\beta}}\bar{Q}_{2}^{+}\rangle-\sec c_{2}\epsilon_{\dot{\alpha}\dot{\beta}}\frac{\sec x_{2}^{-}x_{1}^{-}}{\sec x_{2}^{+}x_{1}^{+}}|\phi^{-}Q_{1}^{+}\tilde{\psi}_{2}^{\dot{\beta}}\rangle)\\
 & = & -\epsilon_{\dot{\alpha}\dot{\beta}}(c_{1}\frac{x_{1}^{-}}{x_{1}^{+}}K_{12}+\sec c_{2}\frac{\sec x_{2}^{-}x_{1}^{-}}{\sec x_{2}^{+}x_{1}^{+}}G_{12})|\phi^{-}\psi_{2}^{\dot{\beta}}\bar{Q}_{1}^{+}\rangle\\
 & - & \epsilon_{\dot{\alpha}\dot{\beta}}(c_{1}\frac{x_{1}^{-}}{x_{1}^{+}}L_{12}+\sec c_{2}\frac{\sec x_{2}^{-}x_{1}^{-}}{\sec x_{2}^{+}x_{1}^{+}}H_{12})|\phi^{-}Q_{1}^{+}\tilde{\psi}_{2}^{\dot{\beta}}\rangle\, .\end{eqnarray*}
This gives another pair of constraints on the coefficients, \begin{eqnarray}
A_{12} & = & \frac{c_{1}}{c_{2}}\frac{x_{2}^{+}}{x_{2}^{-}}\frac{x_{1}^{-}}{x_{1}^{+}}K_{12}+\frac{\sec c_{2}}{c_{2}}\frac{x_{2}^{+}}{x_{2}^{-}}\frac{\sec x_{2}^{-}x_{1}^{-}}{\sec x_{2}^{+}x_{1}^{+}}G_{12}\\
A_{12} & = & \frac{c_{1}}{\sec c_{1}}\frac{x_{2}^{+}}{x_{2}^{-}}\frac{\sec x_{1}^{+}}{\sec x_{1}^{-}}\frac{x_{1}^{-}}{x_{1}^{+}}L_{12}+\frac{\sec c_{2}}{\sec c_{1}}\frac{x_{2}^{+}}{x_{2}^{-}}\frac{\sec x_{1}^{+}}{\sec x_{1}^{-}}\frac{\sec x_{2}^{-}x_{1}^{-}}{\sec x_{2}^{+}x_{1}^{+}}H_{12}\end{eqnarray}

\subsubsection*{Bosonic singlet: Determining $B,\, C$}

To evaluate the $B$ and $C$ matrix elements, we have to study the scattering
of two bosons of opposite spins. Requiring $[\gen Q_{+}^{+},S]=0$
is sufficient to determine them. From
\begin{eqnarray*}
{\cal \gen Q}_{+}^{+}S_{12}|Q_{1}^{+}\bar{Q}_{2}^{-}\rangle & = & {\cal \gen Q}_{+}^{+}[(\frac{1}{2}A_{12}+\frac{1}{2}B_{12})|Q_{2}^{+}\bar{Q}_{1}^{-}\rangle+(\frac{1}{2}A_{12}-\frac{1}{2}B_{12})|Q_{2}^{-}\bar{Q}_{1}^{+}\rangle\\
 & + & \frac{1}{2}C_{12}(|\psi_{2}^{+}\tilde{\psi}_{1}^{-}\phi^{-}\rangle-|\psi_{2}^{-}\tilde{\psi}_{1}^{+}\phi^{-}\rangle)]\\
 & = & a_{2}(\frac{1}{2}A_{12}+\frac{1}{2}B_{12})|\psi_{2}^{+}\bar{Q}_{1}^{-}\rangle+\sec a_{1}(\frac{1}{2}A_{12}-\frac{1}{2}B_{12})|Q_{2}^{-}\tilde{\psi}_{1}^{+}\rangle\\
 & - & \sec b_{1}\frac{1}{2}C_{12}|\psi_{2}^{+}\bar{Q}_{1}^{-}\phi^{+}\phi^{-}\rangle-b_{2}\frac{1}{2}C_{12}|Q_{2}^{-}\check{\phi}^{+}\tilde{\psi}_{1}^{+}\phi^{-}\rangle\\
 & = & a_{2}(\frac{1}{2}A_{12}+\frac{1}{2}B_{12})|\psi_{2}^{+}\bar{Q}_{1}^{-}\rangle+\sec a_{1}(\frac{1}{2}A_{12}-\frac{1}{2}B_{12})|Q_{2}^{-}\tilde{\psi}_{1}^{+}\rangle\\
 & - & \sec b_{1}\frac{1}{2}C_{12}|\psi_{2}^{+}\bar{Q}_{1}^{-}\rangle-b_{2}\frac{1}{2}C_{12}\frac{\sec x_{1}^{-}}{\sec x_{1}^{+}}|Q_{2}^{-}\tilde{\psi}_{1}^{+}\rangle\\
S_{12}{\cal \gen Q}_{+}^{+}|Q_{1}^{+}\bar{Q}_{2}^{-}\rangle & = & S_{12}a_{1}|\psi_{1}^{+}\bar{Q}_{2}^{-}\rangle\\
 & = & a_{1}[K_{12}|\psi_{2}^{+}\bar{Q}_{1}^{-}\rangle+L_{12}|Q_{2}^{-}\tilde{\psi}_{1}^{+}\rangle]\end{eqnarray*}
we find
\begin{eqnarray}
a_{2}\frac{A_{12}+B_{12}}{2}-\sec b_{1}\frac{C_{12}}{2} & = & a_{1}K_{12}\\
\sec a_{1}\frac{A_{12}-B_{12}}{2}-b_{2}\frac{\sec x_{1}^{-}}{\sec x_{1}^{+}}\frac{C_{12}}{2} & = & a_{1}L_{12}\,.\end{eqnarray}
We now turn to the scattering of fermions.

\subsubsection*{$SU(1|2)$ Subsector: Determining $D$}

As before, we first focus on the $SU(1_{\dot{\alpha}}|2_{I})$
sector and consider the scattering of  two fermions in the triplet of $SU(2)_{\dot{\alpha}}$.
This sector will enable us to determine $D$. We look at the condition
$[\gen S_{+}^{I},S]=0$. From \begin{eqnarray*}
\gen S_{+}^{I}S_{12}|\psi_{1}^{+}\tilde{\psi}_{2}^{+}\rangle & = & \gen S_{+}^{I}D_{12}|\psi_{2}^{+}\tilde{\psi}_{1}^{+}\rangle\\
 & = & D_{12}d_{2}|Q_{2}^{I}\tilde{\psi}_{1}^{+}\rangle-D_{12}\sec d_{1}|\psi_{2}^{+}\bar{Q}_{1}^{I}\rangle\\
S_{12}\gen S_{+}^{I}|\psi_{1}^{+}\psi_{2}^{+}\rangle & = & S_{12}(d_{1}|Q_{1}^{I}\tilde{\psi}_{2}^{+}\rangle-\sec d_{2}|\psi_{1}^{+}\bar{Q}_{2}^{I}\rangle)\\
 & = & (d_{1}H_{12}-\sec d_{2}L_{12})|Q_{2}^{I}\tilde{\psi}_{1}^{+}\rangle+(d_{1}G_{12}-\sec d_{2}K_{12})|\psi_{2}^{+}\bar{Q}_{1}^{I}\rangle\end{eqnarray*}
we find \begin{eqnarray}
D_{12} & = & \frac{d_{1}}{d_{2}}H_{12}-\frac{\sec d_{2}}{d_{2}}L_{12}\\
D_{12} & = & -\frac{d_{1}}{\sec d_{1}}G_{12}+\frac{\sec d_{2}}{\sec d_{1}}K_{12}.\end{eqnarray}
A consistent solution needs to satisfy both  equations.

\subsubsection*{Fermionic singlet: Determining $E,F$ }

To determine the remaining coefficients $E$ and $F$, we scatter
two fermions of opposite spins. It is sufficient to require $[\gen S_{+}^{+},S]=0$.
From
\begin{eqnarray*}
\gen S_{+}^{+}S_{12}|\psi_{1}^{+}\tilde{\psi}_{2}^{-}\rangle & = & \gen S_{+}^{+}[(\frac{1}{2}D_{12}+\frac{1}{2}E_{12})|\psi_{2}^{+}\tilde{\psi}_{1}^{-}\rangle+(\frac{1}{2}D_{12}-\frac{1}{2}E_{12})|\psi_{2}^{-}\tilde{\psi}_{1}^{+}\rangle\\
 & + & \frac{1}{2}F_{12}(|Q_{2}^{+}\bar{Q}_{1}^{-}\phi^{+}\rangle-|Q_{2}^{-}\bar{Q}_{1}^{+}\phi^{+}\rangle)]\\
 & = & d_{2}(\frac{1}{2}D_{12}+\frac{1}{2}E_{12})|Q_{2}^{+}\tilde{\psi}_{1}^{-}\rangle-\sec d_{1}(\frac{1}{2}D_{12}-\frac{1}{2}E_{12})|\psi_{2}^{-}\bar{Q}_{1}^{+}\rangle\\
 & + & \frac{1}{2}F_{12}(\sec c_{1}|Q_{2}^{+}\tilde{\psi}_{1}^{-}\phi^{-}\phi^{+}\rangle-c_{2}|\psi_{2}^{-}\check{\phi}^{-}\bar{Q}_{1}^{+}\phi^{+}\rangle)\\
 & = & \frac{1}{2}(d_{2}D_{12}+d_{2}E_{12}+\sec c_{1}F_{12})|Q_{2}^{+}\tilde{\psi}_{1}^{-}\rangle\\
 & + & \frac{1}{2}(-\sec d_{1}D_{12}+\sec d_{1}E_{12}-c_{2}\frac{\sec x_{1}^{+}}{\sec x_{1}^{-}}F_{12})|\psi_{2}^{-}\bar{Q}_{1}^{+}\rangle\\
S_{12}\gen S_{+}^{+}|\psi_{1}^{+}\tilde{\psi}_{2}^{-}\rangle & = & S_{12}d_{1}|Q_{1}^{+}\tilde{\psi}_{2}^{-}\rangle\\
 & = & d_{1}(G_{12}|\psi_{2}^{-}\bar{Q}_{1}^{+}\rangle+H_{12}|Q_{2}^{+}\tilde{\psi}_{1}^{-}\rangle)\end{eqnarray*}
we find \begin{eqnarray}
d_{2}\frac{D_{12}+E_{12}}{2}+\sec c_{1}\frac{F_{12}}{2} & = & d_{1}H_{12}\\
-\sec d_{1}\frac{D_{12}-E_{12}}{2}-c_{2}\frac{\sec x_{1}^{+}}{\sec x_{1}^{-}}\frac{F_{12}}{2} & = & d_{1}G_{12}.\end{eqnarray}

In summary, a sufficient set of linear equations that determine all the coefficients is:

\begin{eqnarray}\label{eqset}
A_{12} & = & \frac{a_{1}}{a_{2}}K_{12}+\frac{\sec a_{2}}{a_{2}}G_{12} \\
A_{12} & = & \frac{a_{1}}{\sec a_{1}}L_{12}+\frac{\sec a_{2}}{\sec a_{1}}H_{12}.\nonumber \\
A_{12} & = & \frac{c_{1}}{c_{2}}\frac{x_{2}^{+}}{x_{2}^{-}}\frac{x_{1}^{-}}{x_{1}^{+}}K_{12}+\frac{\sec c_{2}}{c_{2}}\frac{x_{2}^{+}}{x_{2}^{-}}\frac{\sec x_{2}^{-}x_{1}^{-}}{\sec x_{2}^{+}x_{1}^{+}}G_{12}\nonumber \\
A_{12} & = & \frac{c_{1}}{\sec c_{1}}\frac{x_{2}^{+}}{x_{2}^{-}}\frac{\sec x_{1}^{+}}{\sec x_{1}^{-}}\frac{x_{1}^{-}}{x_{1}^{+}}L_{12}+\frac{\sec c_{2}}{\sec c_{1}}\frac{x_{2}^{+}}{x_{2}^{-}}\frac{\sec x_{1}^{+}}{\sec x_{1}^{-}}\frac{\sec x_{2}^{-}x_{1}^{-}}{\sec x_{2}^{+}x_{1}^{+}}H_{12}\nonumber \\
a_{1}K_{12} & = & \frac{1}{2}a_{2}(A_{12}+B_{12})-\frac{1}{2}\tilde{b}_{1}C_{12}\nonumber \\
a_{1}L_{12} & = & \frac{1}{2}\sec a_{1}(A_{12}-B_{12})-\frac{1}{2}b_{2}\frac{\sec x_{1}^{-}}{\sec x_{1}^{+}}C_{12}\nonumber \\
D_{12} & = & \frac{d_{1}}{d_{2}}H_{12}-\frac{\sec d_{2}}{d_{2}}L_{12}\nonumber \\
D_{12} & = & -\frac{d_{1}}{\sec d_{1}}G_{12}+\frac{\sec d_{2}}{\sec d_{1}}K_{12}\nonumber \\
d_{1}H_{12} & = & \frac{1}{2}d_{2}(D_{12}+E_{12})+\frac{1}{2}\sec c_{1}F_{12}\nonumber \\
d_{1}G_{12} & = & -\frac{1}{2}\sec d_{1}(D_{12}-E_{12})-\frac{1}{2}c_{2}\frac{\sec x_{1}^{+}}{\sec x_{1}^{-}}F_{12}\,. \nonumber
\end{eqnarray}

\bibliographystyle{JHEP}
\bibliography{Orbifoldbib}

\providecommand{\href}[2]{#2}\begingroup\raggedright\begin{thebibliography}{10}

\bibitem{Minahan:2002ve}
J.~A. Minahan and K.~Zarembo, {\it {The Bethe-ansatz for N = 4 super
  Yang-Mills}},  {\em JHEP} {\bf 03} (2003) 013,
  [\href{http://xxx.lanl.gov/abs/hep-th/0212208}{{\tt hep-th/0212208}}].

\bibitem{Beisert:2003yb}
N.~Beisert and M.~Staudacher, {\it {The N=4 SYM Integrable Super Spin Chain}},
  {\em Nucl. Phys.} {\bf B670} (2003) 439--463,
  [\href{http://xxx.lanl.gov/abs/hep-th/0307042}{{\tt hep-th/0307042}}].

\bibitem{Beisert:2003tq}
N.~Beisert, C.~Kristjansen, and M.~Staudacher, {\it {The dilatation operator of
  N = 4 super Yang-Mills theory}},  {\em Nucl. Phys.} {\bf B664} (2003)
  131--184, [\href{http://xxx.lanl.gov/abs/hep-th/0303060}{{\tt
  hep-th/0303060}}].

\bibitem{Staudacher:2004tk}
M.~Staudacher, {\it {The factorized S-matrix of CFT/AdS}},  {\em JHEP} {\bf 05}
  (2005) 054, [\href{http://xxx.lanl.gov/abs/hep-th/0412188}{{\tt
  hep-th/0412188}}].

\bibitem{Beisert:2005tm}
N.~Beisert, {\it {The su(2|2) dynamic S-matrix}},  {\em Adv. Theor. Math.
  Phys.} {\bf 12} (2008) 945,
  [\href{http://xxx.lanl.gov/abs/hep-th/0511082}{{\tt hep-th/0511082}}].

\bibitem{Beisert:2005fw}
N.~Beisert and M.~Staudacher, {\it {Long-range PSU(2,2|4) Bethe ansaetze for
  gauge theory and strings}},  {\em Nucl. Phys.} {\bf B727} (2005) 1--62,
  [\href{http://xxx.lanl.gov/abs/hep-th/0504190}{{\tt hep-th/0504190}}].

\bibitem{Janik:2006dc}
R.~A. Janik, {\it {The AdS(5) x S**5 superstring worldsheet S-matrix and
  crossing symmetry}},  {\em Phys. Rev.} {\bf D73} (2006) 086006,
  [\href{http://xxx.lanl.gov/abs/hep-th/0603038}{{\tt hep-th/0603038}}].

\bibitem{Beisert:2006ez}
N.~Beisert, B.~Eden, and M.~Staudacher, {\it {Transcendentality and crossing}},
   {\em J. Stat. Mech.} {\bf 0701} (2007) P021,
  [\href{http://xxx.lanl.gov/abs/hep-th/0610251}{{\tt hep-th/0610251}}].

\bibitem{Arutyunov:2006iu}
G.~Arutyunov and S.~Frolov, {\it {On AdS(5) x S**5 string S-matrix}},  {\em
  Phys. Lett.} {\bf B639} (2006) 378--382,
  [\href{http://xxx.lanl.gov/abs/hep-th/0604043}{{\tt hep-th/0604043}}].

\bibitem{Beisert:2006ib}
N.~Beisert, R.~Hernandez, and E.~Lopez, {\it {A crossing-symmetric phase for
  AdS(5) x S**5 strings}},  {\em JHEP} {\bf 11} (2006) 070,
  [\href{http://xxx.lanl.gov/abs/hep-th/0609044}{{\tt hep-th/0609044}}].

\bibitem{Agarwal:2010tx}
A.~Agarwal and D.~Young, {\it {SU(2|2) for Theories with Sixteen Supercharges
  at Weak and Strong Coupling}},  {\em Phys. Rev.} {\bf D82} (2010) 045024,
  [\href{http://xxx.lanl.gov/abs/1003.5547}{{\tt 1003.5547}}].

\bibitem{Beisert:2005he}
N.~Beisert and R.~Roiban, {\it {The Bethe ansatz for Z(S) orbifolds of N = 4
  super Yang- Mills theory}},  {\em JHEP} {\bf 11} (2005) 037,
  [\href{http://xxx.lanl.gov/abs/hep-th/0510209}{{\tt hep-th/0510209}}].

\bibitem{Solovyov:2007pw}
A.~Solovyov, {\it {Bethe Ansatz Equations for General Orbifolds of N=4 SYM}},
  {\em JHEP} {\bf 04} (2008) 013,
  [\href{http://xxx.lanl.gov/abs/0711.1697}{{\tt 0711.1697}}].

\bibitem{Gadde:2010zi}
A.~Gadde, E.~Pomoni, and L.~Rastelli, {\it {Spin Chains in N=2 Superconformal
  Theories: from the $Z_2$ Quiver to Superconformal QCD}},
  \href{http://xxx.lanl.gov/abs/1006.0015}{{\tt 1006.0015}}.

\bibitem{fullH}
P.~Liendo, E.~Pomoni, and L.~Rastelli, {\it {To appear}}, .

\bibitem{Hofman:2006xt}
D.~M. Hofman and J.~M. Maldacena, {\it {Giant magnons}},  {\em J. Phys.} {\bf
  A39} (2006) 13095--13118, [\href{http://xxx.lanl.gov/abs/hep-th/0604135}{{\tt
  hep-th/0604135}}].

\bibitem{Kachru:1998ys}
S.~Kachru and E.~Silverstein, {\it {4d conformal theories and strings on
  orbifolds}},  {\em Phys. Rev. Lett.} {\bf 80} (1998) 4855--4858,
  [\href{http://xxx.lanl.gov/abs/hep-th/9802183}{{\tt hep-th/9802183}}].

\bibitem{Klebanov:1999rd}
I.~R. Klebanov and N.~A. Nekrasov, {\it {Gravity duals of fractional branes and
  logarithmic RG flow}},  {\em Nucl. Phys.} {\bf B574} (2000) 263--274,
  [\href{http://xxx.lanl.gov/abs/hep-th/9911096}{{\tt hep-th/9911096}}].

\bibitem{Berenstein:2005jq}
D.~Berenstein, D.~H. Correa, and S.~E. Vazquez, {\it {All loop BMN state
  energies from matrices}},  {\em JHEP} {\bf 02} (2006) 048,
  [\href{http://xxx.lanl.gov/abs/hep-th/0509015}{{\tt hep-th/0509015}}].

\bibitem{Berenstein:2002jq}
D.~E. Berenstein, J.~M. Maldacena, and H.~S. Nastase, {\it {Strings in flat
  space and pp waves from N = 4 super Yang Mills}},  {\em JHEP} {\bf 04} (2002)
  013, [\href{http://xxx.lanl.gov/abs/hep-th/0202021}{{\tt hep-th/0202021}}].

\bibitem{Santambrogio:2002sb}
A.~Santambrogio and D.~Zanon, {\it {Exact anomalous dimensions of N = 4
  Yang-Mills operators with large R charge}},  {\em Phys. Lett.} {\bf B545}
  (2002) 425--429, [\href{http://xxx.lanl.gov/abs/hep-th/0206079}{{\tt
  hep-th/0206079}}].

\bibitem{Aharony:2008ug}
O.~Aharony, O.~Bergman, D.~L. Jafferis, and J.~Maldacena, {\it {N=6
  superconformal Chern-Simons-matter theories, M2-branes and their gravity
  duals}},  {\em JHEP} {\bf 10} (2008) 091,
  [\href{http://xxx.lanl.gov/abs/0806.1218}{{\tt 0806.1218}}].

\bibitem{Nishioka:2008gz}
T.~Nishioka and T.~Takayanagi, {\it {On Type IIA Penrose Limit and N=6
  Chern-Simons Theories}},  {\em JHEP} {\bf 08} (2008) 001,
  [\href{http://xxx.lanl.gov/abs/0806.3391}{{\tt 0806.3391}}].

\bibitem{Gaiotto:2008cg}
D.~Gaiotto, S.~Giombi, and X.~Yin, {\it {Spin Chains in N=6 Superconformal
  Chern-Simons-Matter Theory}},  {\em JHEP} {\bf 04} (2009) 066,
  [\href{http://xxx.lanl.gov/abs/0806.4589}{{\tt 0806.4589}}].

\bibitem{Grignani:2008is}
G.~Grignani, T.~Harmark, and M.~Orselli, {\it {The SU(2) x SU(2) sector in the
  string dual of N=6 superconformal Chern-Simons theory}},  {\em Nucl. Phys.}
  {\bf B810} (2009) 115--134, [\href{http://xxx.lanl.gov/abs/0806.4959}{{\tt
  0806.4959}}].

\bibitem{Bak:2009mq}
D.~Bak, H.~Min, and S.-J. Rey, {\it {Generalized Dynamical Spin Chain and
  4-Loop Integrability in N=6 Superconformal Chern-Simons Theory}},  {\em Nucl.
  Phys.} {\bf B827} (2010) 381--405,
  [\href{http://xxx.lanl.gov/abs/0904.4677}{{\tt 0904.4677}}].

\bibitem{Berenstein:2009qd}
D.~Berenstein and D.~Trancanelli, {\it {S-duality and the giant magnon
  dispersion relation}},  \href{http://xxx.lanl.gov/abs/0904.0444}{{\tt
  0904.0444}}.

\bibitem{Gadde:2009dj}
A.~Gadde, E.~Pomoni, and L.~Rastelli, {\it {The Veneziano Limit of ${\cal N}=2$
  Superconformal QCD: Towards the String Dual of ${\cal N}=2$ $SU(N_c)$ SYM
  with $N_f =2 N_c$}},  \href{http://xxx.lanl.gov/abs/0912.4918}{{\tt
  0912.4918}}.

\bibitem{Berenstein:2005aa}
D.~Berenstein, {\it {Large N BPS states and emergent quantum gravity}},  {\em
  JHEP} {\bf 01} (2006) 125,
  [\href{http://xxx.lanl.gov/abs/hep-th/0507203}{{\tt hep-th/0507203}}].

\bibitem{Berenstein:2005ek}
D.~Berenstein and D.~H. Correa, {\it {Emergent geometry from q-deformations of
  N = 4 super Yang- Mills}},  {\em JHEP} {\bf 08} (2006) 006,
  [\href{http://xxx.lanl.gov/abs/hep-th/0511104}{{\tt hep-th/0511104}}].

\bibitem{Berenstein:2006yy}
D.~Berenstein and R.~Cotta, {\it {Aspects of emergent geometry in the AdS/CFT
  context}},  {\em Phys. Rev.} {\bf D74} (2006) 026006,
  [\href{http://xxx.lanl.gov/abs/hep-th/0605220}{{\tt hep-th/0605220}}].

\bibitem{Arutyunov:2006ak}
G.~Arutyunov, S.~Frolov, J.~Plefka, and M.~Zamaklar, {\it {The off-shell
  symmetry algebra of the light-cone AdS(5) x S**5 superstring}},  {\em J.
  Phys.} {\bf A40} (2007) 3583--3606,
  [\href{http://xxx.lanl.gov/abs/hep-th/0609157}{{\tt hep-th/0609157}}].

\bibitem{MagnonBfield}
N.~Bobev, A.~Gadde, and L.~Rastelli, {\it {Work in progress}}, .

\bibitem{Dorey:2006dq}
N.~Dorey, {\it {Magnon bound states and the AdS/CFT correspondence}},  {\em J.
  Phys.} {\bf A39} (2006) 13119--13128,
  [\href{http://xxx.lanl.gov/abs/hep-th/0604175}{{\tt hep-th/0604175}}].

\bibitem{Roiban:2006gs}
R.~Roiban, {\it {Magnon bound-state scattering in gauge and string theory}},
  {\em JHEP} {\bf 04} (2007) 048,
  [\href{http://xxx.lanl.gov/abs/hep-th/0608049}{{\tt hep-th/0608049}}].

\bibitem{Dorey:2007an}
N.~Dorey and K.~Okamura, {\it {Singularities of the Magnon Boundstate
  S-Matrix}},  {\em JHEP} {\bf 03} (2008) 037,
  [\href{http://xxx.lanl.gov/abs/0712.4068}{{\tt 0712.4068}}].

\end{thebibliography}\endgroup

\end{document}